\documentclass[useAMS,usenatbib]{mn2e}
\input{psfig.tex}

\def\kms{km\,s$^{-1}$}
\def\ene{erg\,s$^{-1}$}

\def\Ha{H$\alpha$}
\def\Hb{H$\beta$}

\def\CaII{Ca\,{\sc ii}}
\def\CII{C\,{\sc ii}} 
\def\OI{O\,{\sc i}}
\def\HII{H\,{\sc ii}}
\def\SiII{Si\,{\sc ii}}
\def\HeI{He\,{\sc i}}
\def\FeII{Fe\,{\sc ii}}
\def\NaID{Na\,{\sc i}~D}
\def\NeI{Ne\,{\sc i}}
\def\aap{A\&A~}
\def\to{$\rightarrow$}

\title[The Broad-Lined Type Ic SN 2003jd]{The Broad-lined Type Ic SN 2003jd\thanks{Based on pbservations at ESO-Paranal, Prog. 074.D-0161A}}

\author[S. Valenti et al.] 
{S. Valenti$^{1,2}$\thanks{E-mail: svalenti@eso.org},
S. Benetti$^{3}$, E. Cappellaro$^{3}$, F. Patat$^{2}$, 
P. Mazzali $^{4,5}$, M. Turatto $^{3}$,
\newauthor
K. Hurley $^6$, K. Maeda $^{4,7}$, 
A. Gal-Yam$^{8}$, R. J. Foley$^{9}$, A. V. Filippenko$^{9}$, 
\newauthor 
A. Pastorello$^{10}$, P. Challis$^{11}$, F. Frontera $^{1,12}$, A. Harutyunyan $^{3}$, M. Iye$^{13}$, 
\newauthor 
K. Kawabata $^{14}$, R.~P. Kirshner$^{11}$, W. Li $^{9}$, Y. M. Lipkin $^{15}$, 
T. Matheson$^{16}$,
\newauthor 
 K. Nomoto$^{17,18}$, E. O. Ofek$^{8}$, Y. Ohyama$^{19}$,  E. Pian$^{5}$, M. Salvo$^{20}$, D. N. Sauer$^{4}$, 
\newauthor 
B. P. Schmidt$^{20}$, A. Soderberg $^{8}$, and L. Zampieri $^{3}$
\\
$^1$ Physics Department, University of Ferrara, I-44100 Ferrara, Italy. \\
$^2$ European Organisation for Astronomical Research in the Southern 
Hemisphere, Karl-Schwarzschild Strasse 2,\\ 
Garching bei M\"{u}nchen, D-85748, Germany.\\
$^3$ INAF - Astronomical Observatory of Padova, I-Padova, Italy.\\
$^4$ Max-Planck Institut f\"{u}r Astrophysik, Karl-Schwarzschild 
Strasse 1, Garching bei M\"{u}nchen, D-85748, Germany.\\
$^{5}$ INAF - Osservatorio Astronomico, Via Tiepolo 11, 34143 Trieste, Italy.\\
$^6$ University of California, Space Sciences Laboratory, Berkeley, 
CA 94720-7450, USA.\\
$^{7}$ Department of Earth Science and Astronomy, College of Arts and 
Science, University of Tokyo, Meguro-ku, Tokyo 153-8902, Japan. \\
$^8$ Division of Physics, Mathematics and Astronomy, California Institute 
of Technology, Pasadena, CA 91125, USA. \\
$^{9}$ Department of Astronomy, University of California, Berkeley,
CA 94720-3411, USA.\\
$^{10}$ Astrophysics Research Centre, School of Mathematics and 
Physics, Queen's University Belfast, Belfast BT7 1NN, United Kingdom.\\
$^{11}$ Harvard-Smithsonian Center for Astrophysics, 60 Garden Street, 
Cambridge, MA 02138, USA.\\
$^{12}$ INAF - Instituto di Astrofisica Spaziale e Fisica Cosmica 
Bologna, via P. Gobetti 101, 40129 Bologna, Italy. \\
$^{13}$ Division of Optical and Infrared Astronomy, NAOJ,  Osawa 
2-21-1, Mitaka, Tokyo 181-8588, Japan. \\
$^{14}$ Hiroshima Astrophysical Science Center, Hiroshima University, 
1-3-1 Kagamiyama, Higashi-Hiroshima, Hiroshima 739-8526, Japan. \\
$^{15}$ School of Physics and Astronomy and Wise Observatory, Tel 
Aviv University, Tel Aviv 69978, Israel.\\
$^{16}$ National Optical Astronomy  Observatory, 950 N. Cherry Ave., 
Tucson, AZ 85719-4933. \\
$^{17}$ Department of Astronomy, School of Science, University of 
Tokyo, Bunkyo-ku, Tokyo 113-0033, Japan. \\
$^{18}$ Research Center for the Early Universe, School of Science, 
University of Tokyo, Bunkyo-ku, Tokyo 113-0033, Japan. \\
$^{19}$ Department of Infrared Astrophysics, ISAS, Japan Aerospace 
Exploration Agency (JAXA), 3-1-1 Yoshinodai, Sagamihara, \\
~~~ Kanagawa 229-8510, Japan. \\
$^{20}$ Research School of Astronomy and Astrophysics, Australian 
National University, Mount Stromlo and Siding Spring Observatories,\\
~~~ Cotter Road, Weston Creek, ACT 2611, Australia.\\ 
}

\begin{document}

\date{Accepted........ Received .........; in original form }

\pagerange{\pageref{firstpage}--\pageref{lastpage}} \pubyear{2002}

\maketitle
\label{firstpage}
\begin{abstract}
The results of a  world-wide coordinated observational campaign on the
broad-lined Type Ic SN~2003jd are presented.  In total, 74 photometric
data  points  and  26   spectra  were  collected  using  11  different
telescopes.   SN  2003jd  is  one  of the  most  luminous  SN~Ic  ever
observed.   A  comparison  with  other  Type  Ic  supernovae  (SNe~Ic)
confirms  that  SN  2003jd  represents an  intermediate  case  between
broad-line events (2002ap, 2006aj),  and highly energetic SNe (1997ef,
1998bw, 2003dh, 2003lw), with an ejected  mass of $M_{ej} = 3.0 \pm 1$
M$_{\odot}$ and  a kinetic energy  of $E_{k}({\rm tot})  = 7_{-2}^{+3}
\times ~10^{51}$~erg.   SN~2003jd is similar to SN~1998bw  in terms of
overall luminosity, but it is closer to SNe 2006aj and 2002ap in terms
of light-curve shape and spectral evolution. The comparison with other
SNe~Ic,  suggests that  the $V$-band  light  curves of  SNe~Ic can  be
partially homogenized by introducing  a time stretch factor.  Finally,
due to the similarity  of SN~2003jd to the SN~2006aj/XRF~060218 event,
we discuss the possible connection of SN 2003jd with a GRB.
\end{abstract}


\section{Introduction}
\label{parintroduction}

In the past decade, the  discovery of the connection of some gamma-ray
bursts (GRBs) with Type Ic supernovae \citep[SNe~Ic; see][for a review
of  supernova classification]{filippenko97}  boosted  interest in  the
study   of   this   type  of   SN   \citep{galama98,stanek03,hjorth03,
malesani04,pian06,soderberg06a,campana06}.

In the  current paradigm there is a  distinction between \emph{normal}
SNe~Ic   of   relatively   low   kinetic   energy   \citep[$E_{51}   =
E_K/(10^{51}~{\rm  erg}) \approx  1$;][]{nomoto01}, of  which SN~1994I
represents  a  prototype,  and  high-energy, broad-lined  events  like
SN~1998bw    \citep[$E_{51}~>10$;][]{iwamoto98,maeda02}    which   are
associated  with  GRBs.   To  understand whether  these  are  separate
subclasses  or  extreme cases  of  a  continuous  distribution, it  is
necessary  to   study  in  detail  intermediate  cases   such  as  the
broad-lined   SN  2002ap   \citep{galyam02,foley03},   which  had   an
intermediate explosion  energy \citep[$E_{51} =  4-10$; ][]{mazzali02}
and  was not  connected with  a GRB,  or the  broad-lined,  low energy
($E_{51}   \approx  2$)   SN   2006aj  associated   with  XRF   060218
\citep{mazzali06a,mazzali07,maeda07}.

Here we  discuss the case  of SN~2003jd, a broad-lined  object showing
clear evidence  of an asymmetric explosion  \citep{mazzali05} but with
no  confirmed GRB  connection  \citep{gcn2434,gcn2439}.  Asymmetry  is
probably a key factor in understanding the diversity of SNe~Ic.

SN 2003jd  was discovered \citep{burket03}  on 2003 Oct. 25  (UT dates
are  used throughout this  paper) with  the Katzman  Automatic Imaging
Telescope  (KAIT)   during  the  Lick   Observatory  Supernova  Search
\citep{filippenko01,filippenko05}. It is located at $\alpha~$
=  $23^h 21^m 03^s.38$  and $\delta$  =-04$\degr 53'  45''.5$ (equinox
J2000),  which is $8.3''$  E and  $7.7''$ S  of the  centre of  the Sb
spiral galaxy MCG-01-59-21 \citep{vandenbergh05}.

\begin{figure}
   \psfig{figure=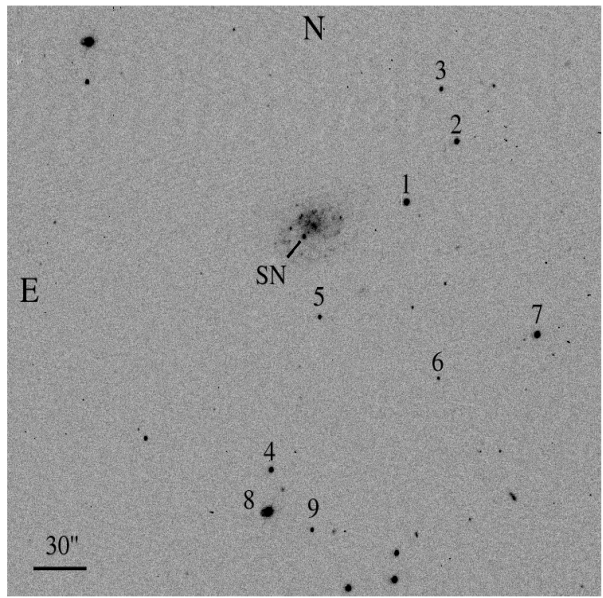,width=8cm,height=8cm}
  \caption{The field of SN 2003jd. $B$-band TNG image (14 Nov. 2003) 
16.4~d after the $B$ maximum.}
  \label{figseqstar}
\end{figure}

The SN was not visible on 16  Oct. in a KAIT unfiltered image (mag $<$
19),  which sets a  tight limit  to the  explosion epoch  ($\leq$ 13~d
before  $B$  maximum). On  28  Oct., SN  2003jd  was  classified as  a
peculiar Type Ic event with  very broad features analogous to those of
SN 1998bw and SN 2002ap \citep{filippenko03}.

The new interest in this kind of event prompted an intensive follow-up
campaign  at  different  observing  sites,  lasting  for  about  three
months. Few  more observations  in the late  nebular phase  were taken
about one year later.

This  paper,  which  presents  and discusses  these  observations,  is
organized as follows. In  \S \ref{parobs} we describe the observations
and  data-reduction  techniques.   We  describe  the  photometric  and
spectroscopic  data   of  SN  2003jd   in  \S  \ref{parphot}   and  \S
\ref{parspe}.  In  \S  \ref{parintrinsic}  we  characterize  the  host
galaxy,  and in  \S \ref{parcomparison}  we compare  SN 2003jd  with a
sample  of  well-studied SNe~Ic:  1994I,  1998bw,  2002ap, 2004aw  and
2006aj. \S \ref{parametribolo} presents  the bolometric light curve of
SN  2003jd,  computed  using   all  available  photometric  data.  The
similarity   with   SN   2006aj/GRB   060218  is   discussed   in   \S
\ref{parmissinggrb},  together  with a  discussion  of  the (lack  of)
evidence  for an  associated  GRB. Finally,  in \S  \ref{parphyparam},
using simple  bolometric light-curve  modelling, we derive  some basic
explosion parameters.

\section{Observations and data reduction}
\label{parobs}

\begin{figure}
   \begin{tabular}{cc}
   \psfig{figure=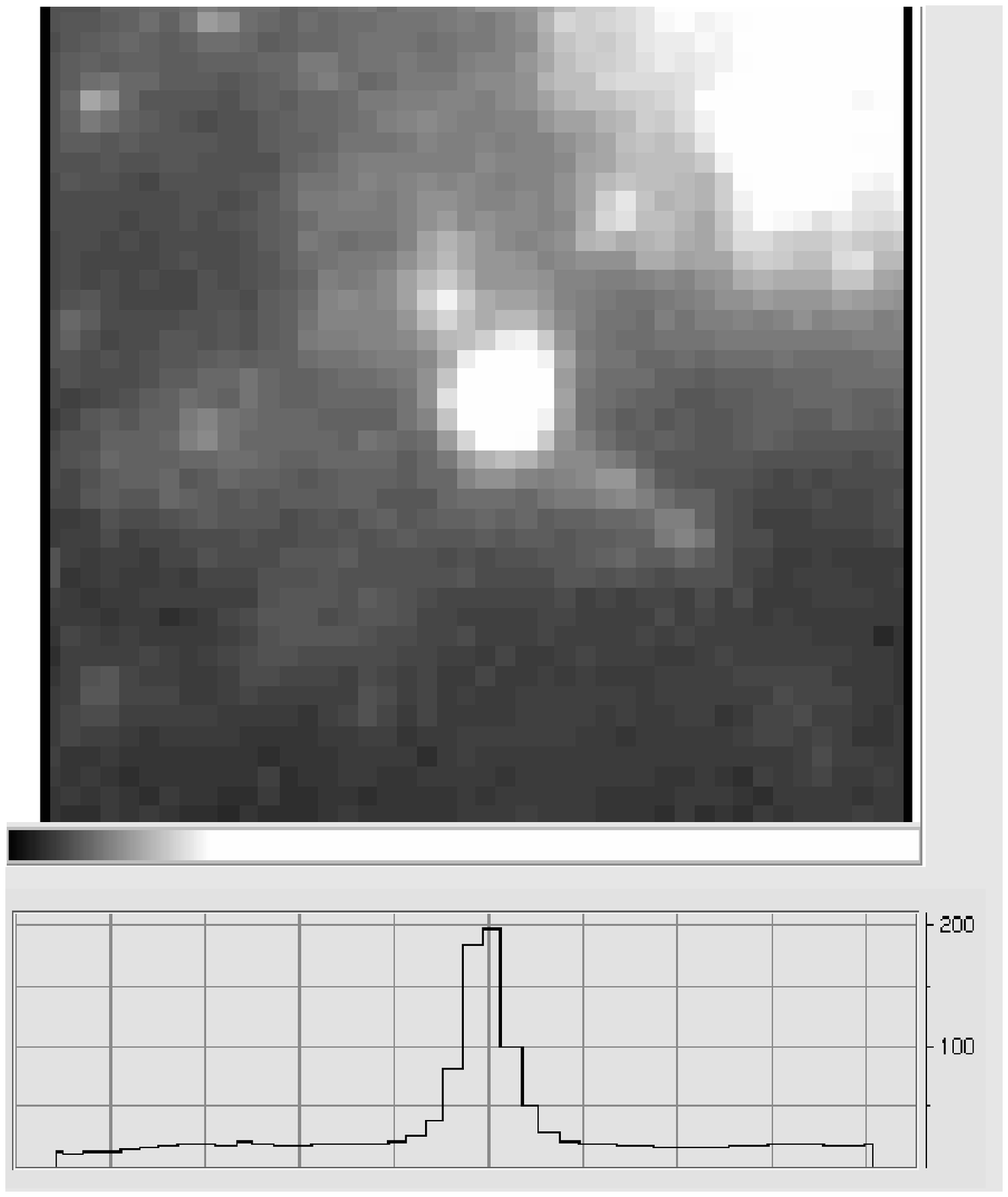,width=4.1cm,height=5.1cm}&
   \psfig{figure=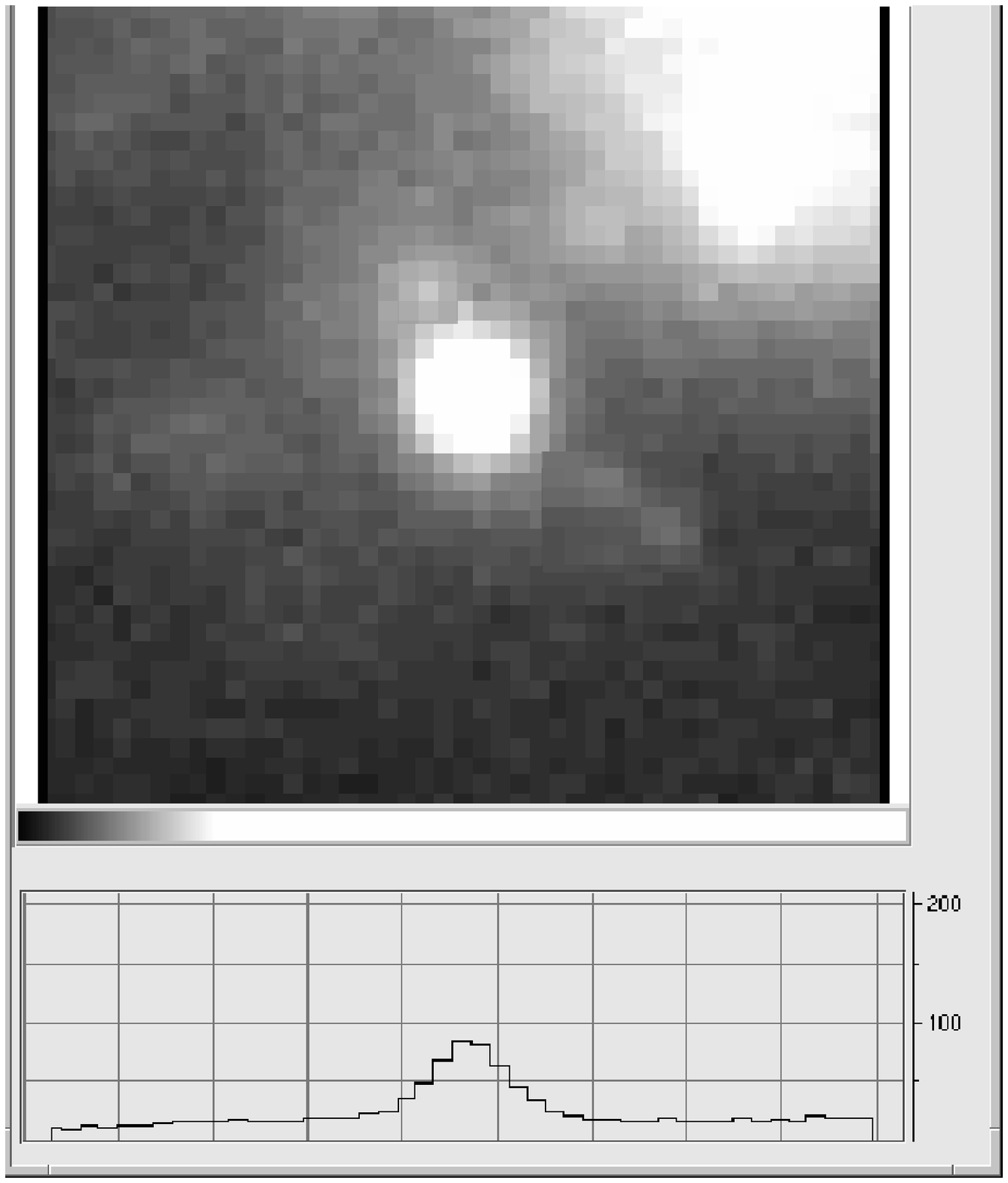,width=4.1cm,height=5.1cm}\\
 \end{tabular}
  \caption{The SN 2003jd region: $\bf{a:}$ $R$-band VLT image 
(2004 Nov. 15, +382~d after $B$ maximum);
$\bf{b:}$ $R$-band Subaru image (2005 Dec. 25, +788~d after 
$B$ maximum) of the star-forming region after SN 2003jd faded 
away. The objects profiles are shown 
 below the images. The profile on the left shows a double 
 component (the broad component is the star-forming region 
and the narrow one is the SN), while on the right side only 
the broad component of the star-forming environment is visible.}
\label{figenv}
\end{figure}

An  extended  spectrophotometric   monitoring  campaign  on  SN~2003jd
started 3~d  before $B$ maximum  and continued through 90~d  after $B$
maximum, using  several telescopes.  Additionally,  three observations
at 10-12  months after explosion  were obtained using the  10~m Keck~I
telescope,  the  Subaru 8.2~m  telescope,  and  the  8.2~m Very  Large
Telescope (VLT).  The  late-time VLT imaging of the  SN obtained under
very  good  seeing  conditions  ($0.5''$) shows  a  broader  component
underlying the stellar point spread function (PSF) of the SN, probably
owing   to    contamination   from   a    star-forming   region   (see
Fig. \ref{figenv}).

All  photometric and  spectroscopic data  were  pre-reduced (trimming,
bias   subtraction,   flat-fielding)   using   IRAF\footnote{IRAF   is
distributed by the National Optical Astronomy Observatories, which are
operated  by  the Association  of  the  Universities  for Research  in
Astronomy, Inc.,  under contract to the  National Science Foundation.}
packages.

In images  where the  SN was bright,  the SN magnitudes  were measured
using   a  PSF-fitting   technique   in  the   SNOOPY\footnote{SNOOPY,
originally presented by Patat (1996),  has been implemented in IRAF by
E. Cappellaro.  The package is  based on DAOPHOT, but optimized for SN
magnitude  measurements.}  package.   At nebular  epochs  (phase $\ge$
70~d), the magnitudes were measured after subtracting a template image
obtained with the Subaru telescope,  788~d after $B$ maximum.  For the
image subtraction  we used  the ISIS package  \citep{alard2000}, which
allows  proper  matching  of  the  PSF  of  the  target  and  template
images.  The  template-subtraction technique  is  an highly  effective
method to  remove background contamination  and allows one  to measure
objects significantly fainter than the background.
%Subtracting the Subaru images (2005 Dec. 25) those 
%from the VLT (2004 Nov. 9), we were able to estimate the SN 
%brightness when it was 2.5 magnitudes fainter than the background.

Even though we didn't use  the template subtraction technique at early
phases, a few  tests of galaxy subtraction at  selected earlier epochs
have  confirmed  that   the  PSF-fitting  technique  gives  consistent
measurements when the SN is bright (deviations $<~3\%$).

Observations of  standard fields \citep{landolt92}  during photometric
nights were  used to  match a local  sequence of reference  stars (see
Fig.  \ref{figseqstar} and Tab.  \ref{tabseqstar}), which were used to
calibrate the observations obtained under non-photometric conditions.

\begin{table}
  \caption{Optical photometry of SN 2003jd reference stars$^a$}
  \label{tabseqstar}
  \begin{tabular}{@{}ccccc@{}}
  \hline
Id & $B$ & $V$ & $R$ & $I$ \\
\hline
\hline
1 & 15.98 (0.01) & 15.38 (0.01) & 14.99 (0.01) & 14.60 (0.01) \\
2 & 16.17 (0.01) & 15.26 (0.01) & 14.73 (0.01) & 14.23 (0.01) \\
3 & 17.86 (0.02) & 17.18 (0.02) & 16.76 (0.02) & 16.43 (0.02) \\
4 & 16.70 (0.02) & 15.98 (0.01) & 15.53 (0.01) & 15.08 (0.01) \\
5 & 18.15 (0.02) & 17.37 (0.02) & 16.87 (0.01) & 16.56 (0.02) \\
6 & 18.23 (0.01) & 16.71 (0.01) & 15.70 (0.01) & 14.44 (0.01) \\
7 & 15.61 (0.02) & 14.99 (0.01) & 14.61 (0.02) & 14.23 (0.02) \\
8 & 13.90 (0.01) & 13.15 (0.01) & 12.75 (0.01) & 12.34 (0.01) \\
9 & 18.35 (0.03) & 17.58 (0.01) & 17.08 (0.02) & 16.85 (0.02) \\
\hline

\end{tabular}
$^a$The uncertainties are the standard deviation of the mean
 of the selected measurements.
%\end{minipage}
\end{table}

Despite   the  moderate  redshift   of  SN~2003jd   ($z=0.0187$),  the
K-correction for  the SN photometry  are not negligible.  We estimated
their values by measuring  the difference between synthetic photometry
in the  observed and rest-frame SN spectra  (see below). K-corrections
between 0.01 and  0.1 mag were measured depending on  the band and the
SN phase.

Spectroscopic observations were pre-reduced  in the same manner as for
imaging, with  the additional  step of fringing  contamination removal
using  the  normalized quartz  standard.   In  the  extraction of  the
spectra we took special  care to minimize the background contamination
subtracting  the spectrum  of  the underlying  star-forming region  as
observed after the SN faded,  two years after the explosion.  Even so,
the  \Ha,  \Hb,  [O\,{\sc  iii}],  [N\,{\sc ii}],  and  [S\,{\sc  ii}]
emission   lines  from  the   \HII~  region   could  not   be  removed
completely.  Fortunately, these  narrow lines  are  well distinguished
from the broad features of the SN.

All spectra  were wavelength and  flux calibrated using arc  lamps and
spectrophotometric  standard-star  spectra  observed during  the  same
night.   In  most  cases  observations  were obtained  with  the  slit
oriented  along  the   parallactic  angle  \citep{filippenko82}.   The
spectra were  also corrected  for atmospheric extinction  and telluric
bands were  removed using standard-star  observations.  The calibrated
spectra  were  compared  with  the  photometry  and,  when  necessary,
multiplied by a constant to correct for slit losses.

\section{Photometry}
\label{parphot}

The $BVRI$ data range from $-3$~d  to 90~d from $B$ maximum with dense
coverage and only a small  temporal gap (3--11~d past $B$ maximum; see
Tab.   \ref{tabopticaldata1}).   We  also   obtained  two   very  late
observations   which  were   used  to   measure  the   slope   of  the
radioactive-decay tail.   In Tab. \ref{tabopticaldata1}  we report the
$BVRI$ magnitude  of SN 2003jd in the  standard Johnson-Cousins system
and the estimated K correction.

\begin{table*}
 \centering
 \begin{minipage}{160mm}
  \caption{Optical photometry of SN 2003jd$^a$}
  \label{tabopticaldata1}
  \begin{tabular}{@{}cccccccc@{}}
  \hline
UT Date & JD $-$ & Phase$^{b}$& $~~~~~~B~~~~~K_{BB}$$^{c}$& $~~~~~~V~~~~~K_{VV}$$^{c}$ & $~~~~~~R~~~~~K_{RR}$$^{c}$ & $~~~~~~I~~~~~K_{II}$$^{c}$ & Source$^{d}$\\
        & 2,400,000\\
\hline
\hline
16/10/03 &  52928.70 &	$-$13. & 	 $-$  	       &	 $-$  	       &	19.0~~~~~~~  08 &	 $-$  	   	&	0,1\\
26/10/03 &  52938.72 &	$-$3.2 & 	16.48 (05)  ~ 02 &       16.24 (23) ~ 05 &	16.15 (02)  08 &	15.91 (03) ~ 00	&	1\\
27/10/03 &  52939.73 &	$-$2.2 & 	16.39 (02)  ~ 01 &	16.12 (24) ~ 05 &	16.02 (02)  08 &	15.80 (02) ~ 00	&	1\\
28/10/03 &  52940.65 &	$-$1.3 & 	16.43 (04)  ~ 01 &	16.10 (04) ~ 05 &	16.01 (03)  08 &	15.81 (04) ~ 00	&	1\\
29/10/03 &  52941.70 &	$-$0.3 & 	16.35 (02)  ~ 00 &	15.98 (02) ~ 04 &	15.92 (02)  07 &	15.68 (02) ~ 00	&	1\\
30/10/03 &  52942.91 &	 0.9  & 	16.33 (03)  ~ 00 &	16.06 (05) ~ 04 &	15.91 (05)  07 &	15.71 (10) ~ 00	&	7	\\
31/10/03 &  52944.38 &	 2.4  & 	16.41 (05) $-$01 &	16.01 (02) ~ 03 &	15.93 (02)  07 &	15.59 (02) ~ 00	&	2	\\
02/11/03 &  52945.69 &	 3.7  & 	16.53 (20) $-$02 &	16.00 (03) ~ 02 &	15.84 (03)  07 &	15.58 (06) ~ 00	&	1	\\
11/11/03 &  52954.74 &	 12.7 & 	17.73 (25) $-$05 &	16.56 (05) $-$01 &	16.19 (05)  06 &	15.71 (06) $-$02	&	1\\
11/11/03 &  52955.34 &	 13.3 & 	17.61 (07) $-$05 &	16.55 (03) $-$02 &	16.20 (04)  06 &	15.82 (02) $-$01	&	2\\	
12/11/03 &  52955.74 &	 13.7 & 	17.82 (10) $-$05 &	16.72 (03) $-$02 &	16.31 (08)  05 &	 $-$ 	   	&	2	\\
12/11/03 &  52956.24 &	 14.2 & 	17.69 (10) $-$05 &	16.60 (06) $-$02 &	16.20 (03)  05 &	15.78 (03) $-$01	&	1\\	
13/11/03 &  52956.74 &	 14.7 & 	17.86 (04) $-$05 &	16.68 (03) $-$02 &	16.24 (11)  05 &	 $-$	   	&	1	\\
13/11/03 &  52957.17 &	 15.1 & 	17.96 (12) $-$05 &	16.75 (02) $-$02 &	16.38 (03)  05 &	15.88 (03) $-$01	&	2\\	
14/11/03 &  52958.17 &	 16.1 & 	17.95 (08) $-$05 &	16.79 (06) $-$02 &	16.41 (07)  05 &	16.03 (06) $-$01	&	2\\
14/11/03 &  52958.44 &	 16.4 & 	18.03 (03) $-$05 &	16.85 (02) $-$02 &	16.42 (02)  05 &	15.91 (02) $-$01	&	9\\	
14/11/03 &  52958.44 &	 16.4 & 	18.03 (03) $-$05 &	16.85 (02) $-$02 &	16.45 (02)  05 &	15.91 (02) $-$01	&	9\\
15/11/03 &  52959.18 &	 17.1 & 	18.10 (10) $-$05 &	16.93 (05) $-$03 &	16.50 (09)  05 &	16.05 (06) $-$02	&	2\\
16/11/03 &  52959.74 &	 17.7 & 	18.02 (06) $-$05 &	16.92 (03) $-$03 &	16.56 (03)  05 &	16.00 (02) $-$02	&	1\\
16/11/03 &  52960.18 &	 18.1 & 	18.26 (12) $-$05 &	17.09 (07) $-$03 &	16.63 (07)  05 &	16.07 (06) $-$02	&	2\\
17/11/03 &  52961.18 &	 19.1 & 	18.17 (05) $-$05 &	17.06 (05) $-$03 &	16.63 (05)  05 &	16.16 (10) $-$02	&	2\\
18/11/03 &  52961.74 &	 19.7 & 	18.32 (12) $-$05 &	17.11 (08) $-$03 &	16.67 (08)  05 &	16.20 (08) $-$02	&	1\\
18/11/03 &  52962.17 &	 20.1 & 	18.27 (06) $-$05 &	17.10 (02) $-$03 &	16.70 (10)  05 &	16.19 (10) $-$02	&	2\\
18/11/03 &  52962.35 &	 20.3 & 	18.39 (06) $-$05 &	17.13 (04) $-$03 &	16.73 (02)  05 &	16.11 (02) $-$02	&	8\\
19/11/03 &  52962.74 &	 20.7 & 	18.32 (09) $-$05 &	17.08 (04) $-$03 &	16.73 (03)  05 &	16.14 (03) $-$02	&	1\\
20/11/03 &  52963.74 &	 21.7 & 	18.33 (08) $-$05 &	17.16 (02) $-$03 &	16.82 (02)  04 &	16.26 (03) $-$02	&	1\\
21/11/03 &  52964.74 &	 22.7 & 	 $-$	       &	17.25 (04) $-$03 &	16.89 (02)  04 &	16.31 (07) $-$03	&	1\\
23/11/03 &  52966.74 &	 24.7 & 	18.68 (23) $-$05 &	17.40 (14) $-$04 &	16.97 (08)  04 &	16.41 (05) $-$03	&	1\\
24/11/03 &  52967.74 &	 25.7 & 	18.52 (16) $-$05 &	17.38 (07) $-$04 &	17.04 (11)  04 &	16.49 (07) $-$03	&	1\\
24/11/03 &  52968.17 &	 26.1 & 	18.73 (08) $-$05 &	17.52 (04) $-$04 &	17.15 (04)  04 &	16.60 (05) $-$03	&	2\\
25/11/03 &  52968.74 &	 26.7 & 	18.71 (07) $-$05 &	17.51 (04) $-$04 &	17.15 (05)  04 &	16.49 (05) $-$03	&	1\\
25/11/03 &  52969.22 &	 27.2 & 	18.71 (13) $-$05 &	17.55 (03) $-$04 &	17.23 (03)  04 &	16.53 (03) $-$03	&	2\\
26/11/03 &  52970.17 &	 28.1 & 	 $-$	       &	17.61 (03) $-$04 &	17.22 (04)  04 &	16.69 (06) $-$04	&	2\\
27/11/03 &  52970.74 &	 28.7 & 	18.69 (05) $-$04 &	17.63 (04) $-$03 &	17.27 (03)  04 &	16.66 (03) $-$04	&	1\\
28/11/03 &  52971.93 &	 29.9 & 	18.68 (06) $-$04 &	17.67 (10) $-$03 &	17.22 (10)  04 &	16.66 (05) $-$04	&	7\\
29/11/03 &  52972.74 &	 30.7 & 	18.82 (28) $-$04 &	17.76 (26) $-$03 &	17.42 (27)  03 &	16.79 (32) $-$04	&	1\\
29/11/03 &  52973.18 &	 31.1 & 	18.91 (14) $-$04 &	17.73 (06) $-$03 &	17.44 (03)  03 &	16.80 (04) $-$04	&	2\\
03/12/03 &  52976.74 &	 34.7 & 	18.85 (10) $-$04 &	17.90 (08) $-$03 &	 $-$	       &	$-$	   	&	1	\\
05/12/03 &  52978.74 &	 36.7 & 	18.95 (08) $-$03 &	17.94 (04) $-$02 &	17.63 (03)  03 &	16.92 (03) $-$05	&	6\\
06/12/03 &  52979.74 &	 37.7 & 	 $-$	         &	18.01 (05) $-$02 &	17.68 (04)  03 &	16.90 (04) $-$06	&	6\\
07/12/03 &  52981.27 &	 39.2 & 	18.93 (22) $-$03 &	18.08 (12) $-$02 &	17.68 (05)  03 &	16.98 (03) $-$06	&	2\\
08/12/03 &  52982.27 &	 40.2 & 	 $-$	       &	18.12 (17) $-$02 &	17.75 (08)  03 &	17.05 (07) $-$06	&	2\\
09/12/03 &  52983.19 &	 41.1 & 	19.13 (21) $-$03 &	18.09 (08) $-$01 &	17.81 (05)  03 &	17.12 (07) $-$06	&	2\\
09/12/03 &  52983.19 &	 41.1 & 	 $-$	       &	18.22 (12) $-$01 &	17.81 (06)  03 &	16.92 (13) $-$06	&	6\\
10/12/03 &  52984.22 &	 42.2 & 	19.05 (12) $-$03 &	18.14 (05) $-$01 &	17.77 (04)  02 &	17.09 (04) $-$06	&	6\\
10/12/03 &  52984.22 &	 42.2 & 	19.20 (29) $-$03 &	18.12 (06) $-$01 &	17.82 (03)  02 &	17.14 (04) $-$06	&	2\\
11/12/03 &  52985.17 &	 43.1 & 	19.12 (10) $-$03 &	18.15 (04) $-$01 &	17.83 (04)  02 &	17.15 (05) $-$07	&	2\\
12/12/03 &  52986.19 &	 44.1 & 	 $-$	       &	18.32 (21) $-$01 &	17.90 (06)  02 &	17.09 (11) $-$07	&	2\\
13/12/03 &  52986.32 &	 44.3 & 	19.11 (21) $-$03 &	18.13 (16) $-$01 &	17.83 (08)  02 &	17.12 (08) $-$07	&	8\\
15/12/03 &  52989.19 &	 47.1 & 	 $-$	       &	18.22 (10)  ~ 00 &	17.94 (04)  02 &	17.31 (08) $-$07	&	2\\
16/12/03 &  52990.18 &	 48.1 & 	 $-$	       &	18.25 (05)  ~ 00 &	18.00 (03)  02 &	17.31 (04) $-$08	&	2\\
17/12/03 &  52991.17 &	 49.1 & 	 $-$	       &	18.28 (06)  ~ 00 &	18.00 (06)  02 &	17.29 (03) $-$08	&	2\\
17/12/03 &  52991.74 &	 49.7 & 	19.19 (10) $-$03 &	18.29 (07)  ~ 00 &	17.97 (05)  02 &	17.28 (07) $-$08	&	1\\
17/12/03 &  52991.74 &	 49.7 & 	19.18 (07) $-$03 &	18.31 (16)  ~ 00 &	18.07 (05)  02 &	17.37 (10) $-$08	&	6\\
18/12/03 &  52992.74 &	 50.7 & 	 $-$	       &	18.40 (08)  ~ 01 &	18.08 (09)  02 &	17.33 (10) $-$08	&	6\\
19/12/03 &  52993.74 &	 51.7 & 	19.26 (17) $-$04 &	18.43 (10)  ~ 01 &	18.16 (25)  02 &	17.30 (16) $-$08	&	1\\
24/12/03 &  52998.28 &	 56.2 & 	 $-$	       &	18.51 (11)  ~ 01 &	18.28 (21)  01 &	17.54 (08) $-$08	&	2\\
27/12/03 &  53000.94 &	 58.9 & 	19.39 (15) $-$04 &	18.55 (15)  ~ 01 &	18.27 (16)  01 &	17.59 (19) $-$09	&	7\\
01/01/04 &  53006.17 &	 64.1 & 	 $-$	       &	18.71 (11)  ~ 01 &	18.41 (12)  00 &	17.72 (08) $-$09	&	2\\
06/01/04 &  53010.74 &	 68.7 & 	$>$19.2	~~~~ $-$05 &	18.73 (12)  ~ 01 &	18.47 (11)  00 &	$>$17.7	~~~~   $-$09	&	1\\
08/01/04 &  53013.21 &	 71.2 & 	 $-$	       &	18.84 (15)  ~ 00 &	18.56 (14)  00 &	17.98 (12) $-$10	&	2\\
\hline	 		 
 \end{tabular}
\end{minipage}		
\end{table*}

\begin{table*}
 \centering
 \begin{minipage}{160mm}
  \begin{tabular}{@{}cccccccc@{}}
   \textbf{Table 2} continued &&&&&&&\\
  \hline

  \hline
UT Date & JD $-$ & Phase$^{b}$& $~~~~~~B~~~~~K_{BB}$$^{c}$& $~~~~~~V~~~~~K_{VV}$$^{c}$ & $~~~~~~R~~~~~K_{RR}$$^{c}$ & $~~~~~~I~~~~~K_{II}$$^{c}$ & Source$^{d}$\\
        & 2,400,000\\
\hline
\hline
09/01/04 &  53013.74&	71.7  &	 $-$	       &	18.77 (15) ~  00 & 	18.58 (25) ~  00 &	 $-$	   	&	1 \\	
11/01/04 &  53016.21&	74.2  &	 $-$	       &	18.89 (05) ~  00 & 	18.60 (05) $-$01 &	18.01 (11) $-$10	& 2 \\	
13/01/04 &  53017.74&	75.7  &	19.73 (17) $-$05 &	18.86 (15) ~  00 & 	18.60 (13) $-$01 &	18.07 (21) $-$09	& 1 \\		
14/01/04 &  53018.74&	76.7  &	19.62 (22) $-$05 &	18.92 (17) $-$01 & 	18.54 (17) $-$01 &	18.11 (23) $-$09	& 1 \\	
15/01/04 &  53020.19&	78.1  &	 $-$	       &	 $-$	         &	18.70 (10) $-$01 &	 $-$	   	&	2 \\		
16/01/04 &  53020.74&	78.7  &	19.60 (24) $-$05 &	18.97 (12) $-$01 & 	18.60 (16) $-$01 &	18.08 (26) $-$09	& 1 \\		
17/01/04 &  53022.19&	80.1  &	 $-$	       &	18.96 (09) $-$01 & 	18.70 (11) $-$01 &	18.31 (16) $-$09	& 2 \\		
17/01/04 &  53022.74&	80.7  &	19.72 (21) $-$04 &	18.80 (11) $-$01 & 	18.66 (16) $-$01 &	18.30 (17) $-$09	& 1 \\		
19/01/04 &  53024.74&	82.7  &	$>$19.3	~~~~   $-$04 &	18.93 (27) $-$01 & 	18.81 (17) $-$01 &	$>$18.1	 ~~~~  $-$09	& 1 \\		
22/01/04 &  53027.18&	85.1  &	 $-$	       &	$-$	       &	$>$18.4	~~~~   $-$01 &	$>$18.5	 ~~~~  $-$09	&2 \\		
26/01/04 &  53031.18&	89.1  &	 $-$	       &	$-$	       &	$>$18.2	~~~~   $-$01 &	$-$  	   	&	2 \\		
11/09/04 &  53259.10&	317.1 &	23.00 (21) ~ 01 &	$-$	       &	21.92 (31) ~  00 &	$-$  	    	&	3 \\		
11/09/04 &  53259.10&	317.1 &	 $-$	       &	23.38 (40) ~ 00 & 	 $-$	         &	$-$  	    	&	3$^e$ \\		
15/11/04 &  53324.02&	382.  &	23.90 (31) ~  01 &	24.38 (35) $-$01 & 	23.61 (50) ~ 00  &	$>$23.0	~~~~~  00	&	4 \\	
\hline
25/12/05  &  53730.1  & 788    &  21.6  & 21.3  & 20.6   & 21.2  & 	3$^f$  	\\
\hline
\end{tabular}
$^a$ The errors (in parentheses, in decimal) are computed taking into account 
both the uncertainty of the PSF fitting of the SN magnitude and 
the uncertainty due to the background contamination 
(computed by the artificial-star experiment).

$^b$Relative to the $B$ maximum (JD = 2,452,942.0). 

$^c$ K correction (in decimal). 

$^d$0 =  IAU Circular 8232. \\  $~~~$ 1  =  0.76~m Katzman  Automatic
 Imaging  Telescope,   Lick  Observatory,  Mt.    Hamilton, CA (USA).\\  
 $~~~$  2  =  1~m Telescope,  Wise  Observatory,  Negev
 desert, Mitzpe  Ramon (Israel).  \\ $~~~$ 3  = 8.2~m  Subaru Telescope,
 National Astr.  Obs. of Japan, Mauna  Kea (Hawaii). \\ $~~~$ 4 = 8.2~m
 ESO-VLT Telescope,  Paranal Observatory, Atacama (Chile). \\  $~~~$ 5 =
 3.5~m Telescopio  Nazionale Galileo,  La Palma, Canary  Isl. (Spain). \\
 $~~~$  6   =  1.5~m  Telescope,  Palomar   Observatory,  Mt.  Palomar,
 CA (USA). \\   $~~~$  7  =  2.3~m   Advanced  Technology  Telescope,
 Siding Spring Observatory,  Coonabarabran (Australia). \\ $~~~$  8 =
 1.82~m Copernico  Telescope, INAF - Osservatorio di  Asiago, Mt. Ekar,
 Asiago (Italy). 

$^e$Synthetic photometry  from spectrum.

$^f$Lower  limit for the magnitude of the star-forming region at  
the position of the SN, measured with aperture photometry in a 
circle of radius $0.5''$.

\end{minipage}
\end{table*}

\begin{figure}
   \psfig{figure=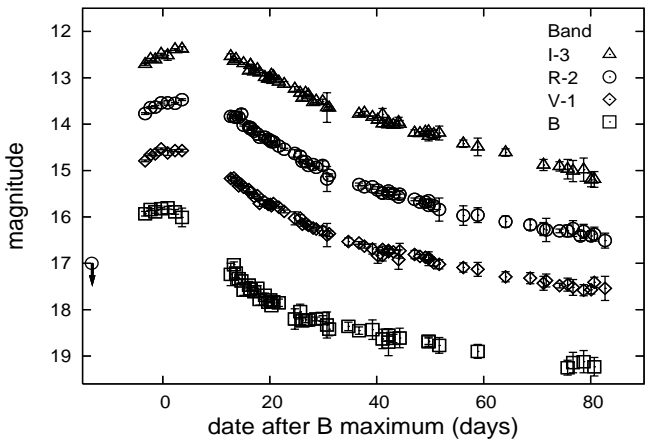,width=9cm,height=6cm,angle=-90}
    \caption{Light curves of SN 2003jd during the photospheric phase, 
ranging from $-3$ to 90~d from $B$ maximum. The error bars are computed 
taking into account both the uncertainty of the PSF fitting of the SN 
magnitude and the uncertainty due to the background contamination 
(computed by an artificial-star experiment).}
   \label{figcurv1}
\end{figure}

The light curves of SN 2003jd in the different bands are shown in Fig.
\ref{figcurv1}.  The data  up to 85~d after $B$  maximum were obtained
mainly through  automated follow-up observations with  the KAIT, Wise,
and Palomar  telescopes.  Unfortunately,  the exposure times  used for
these observations were independent  of the atmospheric conditions and
object brightness.  This  results in a low signal  to noise ratio (S/N
$<15$)  for  the SN  images  near 80~d  past  $B$  maximum.  In  those
exposures where the SN was  below the detection limit, we placed upper
limits (see Tab.\ref{tabopticaldata1}).
%Actually, in a  few exposures, the SN was below the
%detection threshold and we could only determine upper limits 

The maximum  luminosity\footnote{Here and generally  elsewhere in this
paper, the maximum magnitude  is computed via a weighted least-squares
polynomial fit  to the  observations.}  occurred on  2003 Oct.   30 at
$16.34 \pm 0.06$ mag  in the $B$ band, and 2, 3,  and 5~d later in the
$V$,  $R$, and  $I$  bands, respectively  (Tab. \ref{tabparam}).   The
epoch of  explosion is likely  shortly before the  pre-discovery image
taken 13~d  before $B$  maximum.  Indeed, 13~d  is even the  rise time
suggested by  spectral modelling  \citep{sauer07}.  This rise  time is
longer than that  of SN 2006aj \citep[9.5~d as  measured from the date
time of  the GRB  ][]{campana06} and  smaller than in  the case  of SN
1998bw \citep[$\sim$ 15~d][]{mazzali01}.  The light curves are typical
of SNe~Ib/c,  with a  rapid decline after  maximum followed by  a much
slower decline. The transition, which  is not well defined in the case
of SN 2003jd, occurs $\sim$ 30~d after $B$ maximum.

\begin{figure}
   \psfig{figure=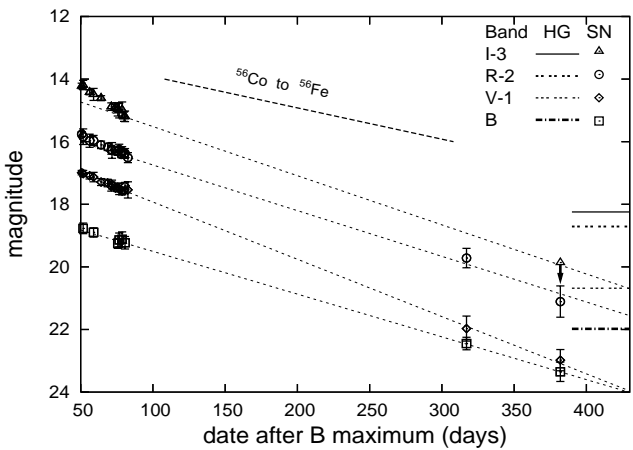,width=8.8cm,height=6cm,angle=-90}
    \caption{Late-time light  curves of SN 2003jd, ranging from
50 to $\sim$400~d past $B$ maximum. The horizontal lines on the right
mark  the magnitude of the background region.}
\label{figcurv2}
\end{figure}

Fig.  \ref{figcurv2}  shows the late-time  light curves of  SN 2003jd,
from 50~d to  383~d after $B$ maximum.  The  short horizontal segments
on the right mark the magnitudes of the background star-forming region
as  measured via  PSF fitting  on the  Subaru images  of 2005  Dec. 25
(seeing $\sim 0.8''$).  The actual background contamination depends on
the  sky conditions  and is,  of course,  larger in  nights  with poor
seeing.

The  late-time luminosity  decline  rates were  computed via  weighted
linear least-squares fits to the observations (see Fig. \ref{figcurv2}
and Tab.  \ref{tabparam}). The  slopes are steeper  in all  bands than
those expected when the energy source is $^{56}$Co~\to~$^{56}$Fe decay
and   the   trapping   of   $\gamma$-rays  is   complete   [0.98   mag
(100~d)$^{-1}$].

Adopting the distance modulus of  the host galaxy MCG-01-59-21 ($\mu =
34.46$ mag;  see \S \ref{parintrinsic})  and assuming $E(B-V)  = 0.14$
mag for the total  interstellar extinction (\S \ref{parintrinsic}), we
estimate that  SN 2003jd reached  an absolute magnitude at  maximum of
$-18.9 \pm  0.3$ in  the $V$  band.  This value  is comparable  to the
magnitude    of   SN    1998bw   ($-19.12    \pm   0.05$)\footnote{See
Tab.  \ref{tabparam2} for  references.}   and much  brighter than  the
magnitudes of SN  2002ap ($-17.37 \pm 0.05$) and  SN 1994I ($-17.5 \pm
0.3$).

\begin{table*}
 \centering
\begin{minipage}{140mm}
\caption{Main parameters for light curves of SN 2003jd}
\label{tabparam}
\begin{tabular}{ccccc}
\hline
Parameter & $B$ & $V$ & $R$ & $I$ \\
\hline
\hline
Date of max (JD $-$2,400,000)         & $52942.0\pm1$ (2003 Oct. 30) & $52944.1\pm1$ & $52944.7\pm1$ & $52947.2\pm1$  \\
Apparent magnitude at max     & $15.75\pm 0.24$ & $15.49\pm 0.18$  & $15.48\pm 0.15$ & $15.36\pm 0.13$ \\
Absolute magnitude at max     & $-18.7 \pm 0.3$ & $-18.9\pm 0.3$ & $-19.0\pm 0.2$ & $-19.1\pm 0.2$ \\
Late-time decline $\gamma$ (mag d$^{-1}$)    & $  0.0141 \pm 0.0013 $   & $  0.0189 \pm 0.0005 $  & $ 0.0149 \pm 0.003  $  & $> 0.016 $ \\
Phase range               &     75--382                &  71--382        &    71--382     & 71--382  \\
\hline
\end{tabular}
\end{minipage}
\end{table*}

\section{Spectroscopy}
\label{parspe}

Spectroscopic monitoring started one  day before $B$ maximum light and
continued until  51~d after  $B$ maximum (see  Tab. \ref{tabseqspec}).
In  addition, using  the Subaru  and Keck  telescopes, we  secured two
spectra at late epochs, 317 and 354~d after $B$ maximum, respectively.

\begin{table*}
 \centering
  \begin{minipage}{140mm}
  \caption{Journal of spectroscopic observations}
  \label{tabseqspec}  
  \begin{tabular}{@{}cccccc@{}}
  \hline  UT Date  & JD  $-$ &  Phase\footnote{Relative to  the B-band
maximum   light   (JD  =   2,452,942.0).}    &   Range  &   Resolution
FWHM\footnote{FWHM     of      night-sky     emission     lines}     &
Equipment\footnote{Ekar   =  1.82~m   Copernico   Telescope,  INAF   -
Osservatorio di Asiago, Mt.  Ekar, Asiago (Italy). \\$~~~$ TNG = 3.5~m
Telescopio  Nazionale  Galileo,  La  Palma,  Canary  Isl.   (Spain).\\
$~~~$Subaru =  8.2~m Subaru Telescope, National Astr.  Obs.  of Japan,
Mauna Kea,  Hawaii (USA).\\$~~~$Keck =  10~m Keck~I Telescope,  W.  M.
Keck  Obs.,  Mauna Kea,  Hawaii  (USA).\\  $~~~$ATT  = 2.3~m  Advanced
Technology   Telescope,  Siding   Spring   Observatory,  Coonabarabran
(Australia).\\  $~~~$FLWO  = 1.5~m  Telescope,  Fred Lawrence  Whipple
Observatory,  Mt. Hopkins, Arizona  (USA). \\  $~~~$Shane =  Shane 3-m
telescope at Lick Observatory, Mt. Hamilton, CA (USA). }\\ & 2,400,000
& (days) &  (\AA) & (\AA) & \\  \hline \hline 2003 Oct 28  & 52940.5 &
$-1$ & 3200--10000 &  7 & Shane+Kast\\ 2003 Oct 28 &  52940.6 & $-1$ &
3750--7550 &  7 & FLWO+FAST+300gpm\\  2003 Oct 29  & 52941.7 & $  0$ &
3700--7550 &  7 & FLWO+FAST+300gpm\\  2003 Oct 30  & 52942.9 &  $+1$ &
3720--7550 &  7 & FLWO+FAST+300gpm\\  2003 Oct 30  & 52943.4 &  $+1$ &
3500--7500 &  25 & Ekar+AFOSC+gm.4\\  2003 Oct 31  & 52944.4 &  $+2$ &
3720--7550 &  7 & FLWO+FAST+300gpm\\  2003 Nov 01  & 52945.0 &  $+3$ &
3720--7550 &  7 & FLWO+FAST+300gpm\\  2003 Nov 04  & 52948.0 &  $+6$ &
3500--7500 &  25 & Ekar+AFOSC+gm.4\\ 2003  Nov 15 & 52959.4  & $+17$ &
3200--10250 & 12  & TNG+DOLORES+gm.LRB,LRR \\ 2003 Nov  18 & 52962.2 &
$+20$ & 3500--10100 & 25 & Ekar+AFOSC+gm.2,4\\ 2003 Nov 19 & 52963.3 &
$+21$ &  3720--7500 & 7 &  FLWO+FAST+300gpm\\ 2003 Nov 20  & 52963.9 &
$+22$ &  3720--7550 & 7 &  FLWO+FAST+300gpm\\ 2003 Nov 20  & 52964.3 &
$+22$ &  3500--7800 & 25 &  Ekar+AFOSC+gm.4\\ 2003 Nov 21  & 52964.7 &
$+23$ &  3720--7550 & 7 &  FLWO+FAST+300gpm\\ 2003 Nov 22  & 52965.5 &
$+24$ &  3720--7550 & 7 &  FLWO+FAST+300gpm\\ 2003 Nov 23  & 52966.5 &
$+25$ & 3200--10000 & 7 & Shane+Kast\\ 2003 Nov 23 & 52966.7 & $+25$ &
3720--7550 &  7 & FLWO+FAST+300gpm\\ 2003  Nov 24 & 52967.8  & $+26$ &
3720--7550 &  7 & FLWO+FAST+300gpm\\ 2003  Nov 27 & 52970.9  & $+29$ &
3720--7550 &  7 & FLWO+FAST+300gpm\\ 2003  Nov 28 & 52971.8  & $+30$ &
3720--7550 &  7 & FLWO+FAST+300gpm\\ 2003  Nov 28 & 52972.3  & $+30$ &
3500--9200 &  6 & ATT+DBS+gm.300B\\  2003 Nov 29  & 52973.5 &  $+31$ &
3200--10000  & 7  &  Shane+Kast\\ 2003  Dec  04 &  52977.7  & $+36$  &
3720--7550 &  7 & FLWO+FAST+300gpm\\ 2003  Dec 19 & 52992.7  & $+51$ &
3600--8820 & 7 & FLWO+FAST+300gpm\\ 2004  Sept 11 & 53259.1 & $+317$ &
3100--9350 &  10 &  Subaru+FOCAS+B300+Y47 \\ 2004  Oct 18 &  53296.1 &
$+354$ & 3100--9350 & 8 & Keck+LRIS \\ 2005 Dec 25 & 53730.2 & $+788$
\footnote{Spectrum  of  the SN  site  26  month  after explosion.}   &
3100--7200 & 10 & Subaru+FOCAS+B300+Y47 \\ \hline
\end{tabular}
\end{minipage}
\end{table*}

\subsection{Photospheric spectra}
\label{parspe2}
\begin{figure*}
 \centering
    \psfig{figure=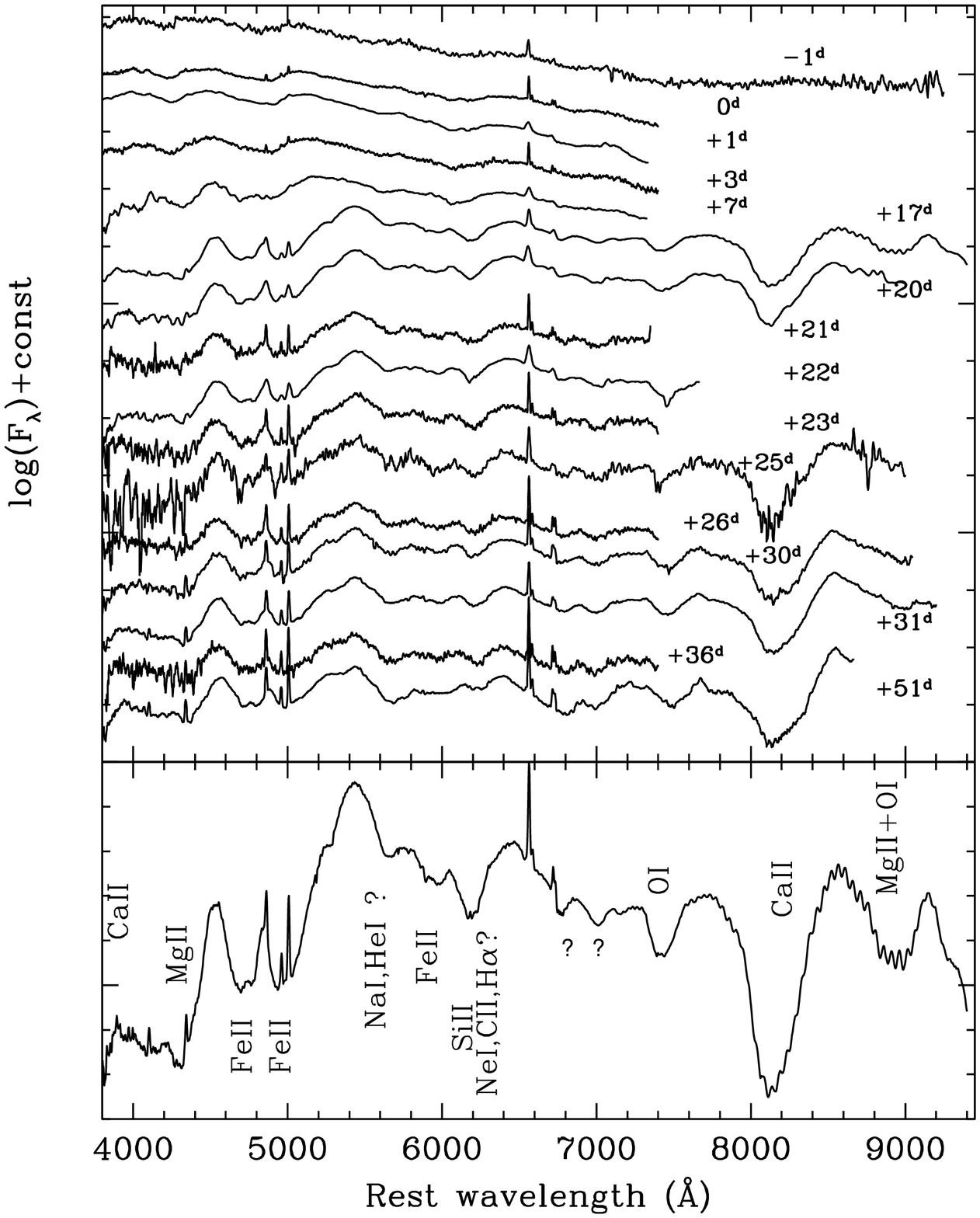,width=15cm,height=18cm}
   \caption{(Top) Spectroscopic evolution of SN 2003jd from $-1$~d to +51~d
with respect to $B$ maximum light.  The wavelength scale is in the supernova
rest frame. The spectra are smoothed over a window of 10~\AA. The narrow 
lines are host-galaxy contamination.
(Bottom) Identification of the main features in the
+17~d spectrum.  The following lines are marked: 
\CaII~ (H\&K, 8498~\AA, 8542~\AA, 8662~\AA); \OI~ (7774~\AA); \FeII~
(4440~\AA, 4555~\AA, 4924~\AA, 5018~\AA, 5170~\AA); \HeI~ (5876~\AA) and 
\NaID~ (5890~\AA, 5896~\AA); Mg\,{\sc ii}~ (4481~\AA). The feature near 
6100~\AA\ is mainly attributed to \SiII~ (6355~\AA), contaminated by 
another feature, probably \NeI, but contamination by  \CII~ or \Ha~
cannot be excluded. The feature near 5650~\AA~ is most likely due to the
\NaID. Contamination by \HeI~ (5876~\AA) cannot be excluded but 
would imply a unusually low velocity for the \HeI~ layer.}
   \label{ddd}
\end{figure*}

The spectra obtained in the first two months after explosion are shown
in Fig. \ref{ddd} (upper panel; wavelengths are in the SN rest frame).
The early-time spectra show broad absorption features similar to those
observed  in SNe 1998bw,  2002ap, and  2006aj. These  features suggest
that a  significant amount of the  gas is moving  at velocities larger
than 15000 \kms.

As the spectra  evolve and the velocity at  which they form decreases,
the spectral  features become less blended  than in SN  1998bw and the
typical features  of normal SNe~Ic  emerge, even if the  spectral time
evolution   remains   slower  than   that   of   SN   1994I  (see   \S
\ref{paragrafostretch}).

The absorption features close  to 5660~\AA\ and 6200~\AA\ are probably
due to \NaID~ and  \SiII, respectively, with possible contamination by
\HeI~ in the \NaID~ feature  and \NeI~ in the \SiII.  Contamination by
\Ha~ or \CII~ to the \SiII~ line can also not be excluded.  The \CaII~
near-infrared  (IR) triplet and  \OI~ at  7770~\AA~ are  prominent and
only marginally blended.

Close to  $\sim$6770~\AA~ and $\sim$7005~\AA, SN 2003jd  shows two not
easily identified  lines. Helium  at $\sim$11,000 \kms\  could explain
these features,  but the complete  absence of the \HeI~  6678~\AA~ and
the low strength  of the feature at $\sim$5650~\AA,  that in this case
should be  identified as \HeI~  at 5876~\AA, make  this identification
unlikely.

\begin{figure}
   \psfig{figure=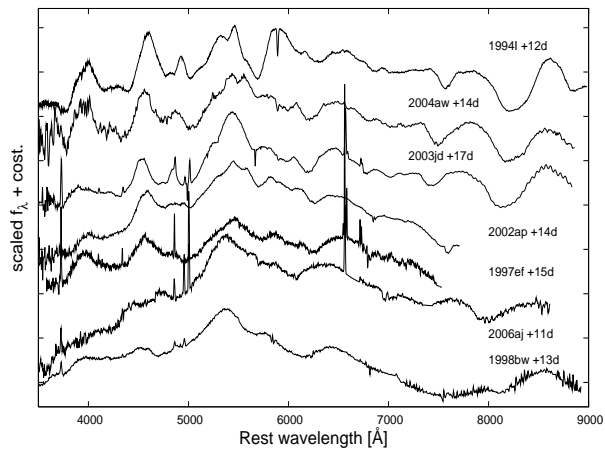,width=8cm,height=6cm,angle=-90}
  \caption{Comparison  spectra of SNe~Ic at $\sim$2 weeks past 
  $B$ maximum.}
  \label{figcomp3}     
\end{figure}

In  Fig. \ref{figcomp3} we  show the  spectrum of  SN 2003jd  $\sim$ 2
weeks  past maximum,  compared with  the  spectra of  other SNe~Ic  at
similar epochs.

The spectra  are arranged  in order of  increasing line width  (top to
bottom), as  exemplified by the iron lines  around 4000--5000~\AA.  In
particular,  the  two  absorptions   with  minima  at  $\sim$4720  and
$\sim$4990~\AA~,  caused  by  the  \FeII~ lines  (4924~\AA,  5018~\AA,
5170~\AA), are clearly visible in the SN 1994I-like SNe~Ic, while they
are partially or completely blended in the other objects.

The increase of the line width is  also visible in the red part of the
spectra, with  feature such as  \OI~ 7774~\AA~ and the  \CaII~ near-IR
triplet.

\subsection{Photospheric velocity}
\label{parvelosity}

Because of  severe line blending in broad-lined  SNe~Ic, estimating of
the photospheric  velocity from direct  measurement of the  minimum of
specific spectral absorption lines is particularly difficult.

We  used  the  \SiII~  line  (6355~\AA)  as the  best  tracer  of  the
photospheric velocity, with the  caveat that other ions may contribute
to the $\sim$6200~\AA~ absorption. In  SN 1994I, for example, a number
of  potential   contaminants  have  been   suggested:  detached  \HeI~
(6678~\AA)    \citep{clocchiatti96},    detached   \CII~    (6580~\AA)
\citep{elmhamdi06},   detached   \Ha~   \citep{branch06},  and   \NeI~
(6402~\AA) \citep{sauer06}.

\begin{figure}
   \psfig{figure=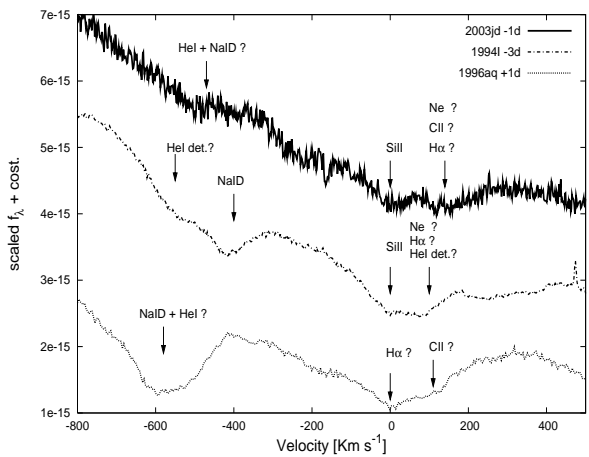,width=8cm,height=6cm,angle=-90}
  \caption{Spectral  comparison   SN 2003jd,  SN 1994I, and  SN
  1996aq  in the  range  between \NaID~  and  \SiII. The zero point of the 
  abscissa is placed at the blue component of  the line at $\sim$6200~\AA~ 
  (rest frame). All of the spectra show a double-dipped or relatively 
  flat-bottomed profile, probably due to more than one component. The 
  line identifications were proposed by Clocchiatti et al. (1996), 
  Elmhamdi et al. (2006), Branch et al. (1996), and Sauer et al. (2006) 
  for SN 1994I, and by Elmhamdi et al. (2006) for type Ib/c SN 1996aq.}
  \label{figsilicumline}     
\end{figure}

In  Fig.  ~\ref{figsilicumline}, we  show a  comparison of  the \SiII~
spectral  region for  three SNe~Ic  (1994I, 2003jd,  and  1996aq), all
showing  a second  component  in the  $\sim$6200~\AA~ absorption  line
indicative of line blending.

As in the case of SN 1994I, we assume that for SN 2003jd only the blue
component is due  to \SiII, whereas the red wing  is dominated by some
other lines. With this  assumption we estimate a photospheric velocity
of $\sim$14,200  \kms\ at $B$  maximum and $\sim$7900 \kms\  one month
later.   These values  are  in good  agreement  with the  photospheric
velocity  deduced from  spectral modelling  \citep[$v  \approx$ 13,500
\kms\ and $v \approx$ 8000 \kms, respectively;][]{sauer07}.

The identification  of the second contributor to  this feature remains
dubious: \NeI~  is a possible candidate,  but \CII~ or  \Ha\ cannot be
excluded as suggested for other SNe Ic \citep{branch06}.

\begin{figure}
   \psfig{figure=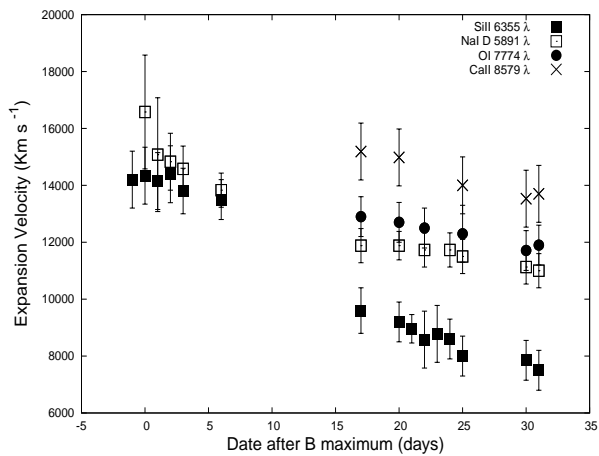,width=8cm,height=6cm,angle=-90}
  \caption{Photospheric velocity for the \NaID~ (5890~\AA, 5896~\AA), \SiII~
  (6355~\AA), \OI~  (7770~\AA), and \CaII~ (8498~\AA, 8542~\AA, 8662~\AA). 
  The  measurements of the \OI~ velocity were  done fitting  a  Gaussian  
  to the absorption minimum. 
  The  fitting of the  \SiII~ line was done  using two Gaussian functions, 
 taking the blue one as the \SiII~ line. 
 The photospheric velocity of \CaII~ and \NaID~ were calculated at 
 the blueshift of the minimum away from an estimated wavelength of 
 8579~\AA~ (\CaII) and 5891~\AA~ (\NaID) via a weighted average of 
 the three and two components of \CaII~ and \NaID, respectively 
 \citep{thomas04}.}
  \label{figvelocita}     
\end{figure}
 
The evolution  with time of  the expansion velocities measured  from a
few   of   the   features   mentioned   above   is   shown   in   Fig.
\ref{figvelocita}.  The \OI~ and  \CaII~ velocities  are significantly
higher than  the \SiII~  velocities, but with  similar slopes.  On the
other hand, the velocity of \NaID~ shows a very rapid decline near $B$
maximum. This is probably not real, but rather due to contamination by
\HeI.
%
%In \S  \ref{parphyparam}, the  photospheric velocity close  to maximum
%brightness  is used along  with other  measurements to  constrain some
%physical  parameters  for SN  2003jd  and  other  SNe~Ic.  The  \SiII~
%absorption, as we argued above, is to be preferred over other lines as
%the photospheric  velocity tracer for SN 2003jd.   The \SiII~ velocity
%is  in agreement  with  the velocity  coming  from spectral  modelling
%\citep{sauer07}.

\subsection{Late-phase spectra and asymmetry}
\label{parlatephase}

\begin{figure}
   \psfig{figure=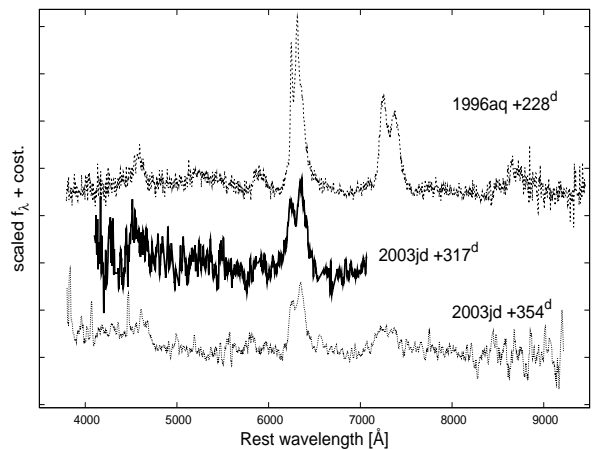,width=8cm,height=6cm,angle=-90}
  \caption{Nebular spectra of SN 2003jd (317~d, 354~d) and 
SN 1996aq (228~d).}
  \label{figlatephase}     
\end{figure}

The  nebular  spectra   \cite[see  also][]{mazzali05}  show  forbidden
emission  lines, in  particular  Mg\,{\sc i}]  and  [\OI], typical  of
SNe~Ic,   but  with   a   peculiar  double-peak   profile  (see   Fig.
\ref{figlatephase}). \cite{mazzali05} interpreted this as evidence for
a strongly  asymmetric explosion with  some of the material  moving at
high  velocity within  a jet-like  structure. The  double-peak profile
suggests an  angle of  $\ge 70^{\circ}$ between  the jet axis  and the
line of sight. This configuration is consistent with the non-detection
of an  associated GRB, because the  GRB would be beamed  along the jet
direction and not towards the observer (see \S \ref{parmissinggrb}).

Recently  more   SN  Ib/c   with  a  double   peak  have   been  found
\citep{maeda07b}.   Here we  show another  SN~Ib/c, namely  SN 1996aq,
showing a double-peak [O\,{\sc i}] emission line similar to that of SN
2003jd (see  Fig. \ref{figlatephase}).   While a detailed  analysis of
the observations  has not yet  been performed, we note  that SN~1996aq
appears to be  a typical SN~Ic with a  moderate photospheric expansion
velocity   (9000   \kms,    as   deduced   from   spectral   modelling
\citep{elmhamdi06})  and  luminosity ($M_V  \approx  -17$ mag).   This
would suggest  that asymmetric explosions are not  unique to energetic
broad-line SNe~Ic.

\section{Environment properties: MGC-01-59-021}
\label{parintrinsic}

There are  no direct  estimates of the  distance to  MGC-01-59-21, the
host  galaxy  of  SN 2003jd,  and  therefore  we  must resort  to  the
recession velocity  and the Hubble  law. However, this method  gives a
good estimate of the distance, albeit with a large uncertainty.

Using the  radial velocity corrected  for Local Group infall  into the
Virgo   Cluster  (5625   \kms;  LEDA)   and  a   Hubble   constant  of
$72~\pm~5$ \kms~  Mpc$^{-1}$ \citep{freedman01,spergel03}, we
derive  a distance  modulus $\mu  = 34.46  \pm 0.20$  mag.   The error
include  the  uncertainty  in   the  Hubble  constant  and  an
uncertainty  of 200  \kms\ for  the radial  velocity  of MGC-01-59-021
owing to peculiar motions.

The  Galactic extinction  in the  direction of  SN 2003jd  reported by
\cite{schlegel98} is $E(B-V)_{Gal} = 0.044$ mag.

The presence in the spectra of a narrow \NaID~ line at the host-galaxy
redshift  with  equivalent width  EW  =  $0.62  \pm 0.02$~\AA~  support
extinction in the host galaxy.  Interstellar gas is usually associated
with dust, and indeed there is a general correlation between \NaID~ EW
and colour excess, although with a large dispersion \citep{turatto03}.

Since  SN 2003jd at  $B$-band maximum  was already  one of  the bluest
SNe~Ic  ever observed  (see  Fig.  \ref{colorcurve}),  we exclude  the
possibility that the relation between  \NaID~ EW and colour excess for
SN     2003jd     can    follow     the     steeper    relation     of
\cite{turatto03}.  Assuming   a  conservative  relation   reported  by
\cite{turatto03}, namely $E(B-V) = 0.16$EW, we estimate $E(B-V)_{host}
\approx 0.10^{+0.10}_{-0.05}$ mag,  where the uncertainty accounts for
the dispersion in the correlation.  The total reddening suffered by SN
2003jd is $E(B-V) \approx 0.14^{+0.10}_{-0.05}$, where the uncertainty
in the Galactic component is negligible.

\begin{table}
\caption{Main parameters for SN 2003jd and its host galaxy}
\label{tabsummary}
\begin{tabular}{ll}
\hline
Parent galaxy           &  MCG-01-59-021     \\
Galaxy type             &  Sb pec$^a$         \\
RA (2000)               &  $23^h 21^m 03^s.38$    \\
Dec (2000)              & $-04^\circ53'45.5''$  \\
Recession velocity      [\kms] & 5625$^b$ \\
Distance modulus ($H_0 = 72$) & $34.46 \pm 0.20$ mag \\
$E(B-V)_{host}$         & $0.10^{+0.10}_{-0.05}$ mag$^c$\\
$E(B-V)_{Gal}$          & 0.044 mag$^d$ \\
Offset from nucleus     &  $8.3''$ E, $7.7''$ N   \\
%                        &                               \\
%Explosion epoch (MJD)   & $52338.1\pm 0.5$ (Mar 05, 2002) \\
%Date of B maximum (MJD) & $52356.0\pm0.5$ (Mar 23, 2002) \\
%Magnitude at max        & $B=14.04\pm 0.10$, $V=13.58\pm 0.10$,\\
%                        & $R=13.49\pm 0.10$, $I=13.52\pm 0.10$ \\
\hline
\end{tabular}\\
$^a$van den Bergh et al. 2005. \\ 
$^b$LEDA, corrected for Local Group infall (208 \kms).\\ 
$^c$Calculated from the equivalent width of the \NaID~ lines 
\citep{turatto03}.\\ 
$^d$Schlegel et al. (1998).
\end{table}

The spectra of the host galaxy  taken at +788~d, long after the fading
of the SN, can be  used to estimate the metallicity and star-formation
rate  (SFR)   of  the  region  were   SN~2003jd  exploded.   Following
\cite{pettini04},  from  the [O\,{\sc  iii}]  and  [N\,{\sc ii}]  line
strengths we derive an oxygen abundance  of $12 + {\rm log}(O/H) = 8.4
\pm 0.1$ dex. A slightly higher  value ($12 + {\rm log}(O/H) = 8.7 \pm
0.1$  dex) is  obtained using  the $R_{23}$  index (([O\,{\sc  iii}] +
[O\,{\sc ii}])/H$\beta$) of \cite{pagel79}.  The two estimates bracket
the measurement by \cite{modjaz07};  they confirm that the metallicity
of the SN~2003jd host is  below average for local galaxies and similar
to that  of the SN~1998bw  host \citep{modjaz07}.  On the  other hand,
MCG-01-59-021 is a luminous galaxy,  $M_{B} = -20.3$ mag (LEDA), which
is  somewhat at  odds with  the medium  to low  metallicity \citep[see
Fig.~5 of ][ and Fig.~1 of Prieto et al 2007]{modjaz07}.

The SFR at  the SN location can be estimated from  the \Ha\ flux, $7.1
\pm   0.5~erg~s^{-1}~cm^{-2}$.    Following   \cite{kennicutt98},   we
estimate a SFR $\approx  0.04$~M$_{\odot}$ yr$^{-1}$, which is in fair
agreement   with  the   value  derived   by  \cite{modjaz07}   (SFR  =
0.07~M$_{\odot}$ yr$^{-1}$).  This SFR  is typical for  \HII~ regions,
and similar  for instance  to that derived  for the environment  of SN
2006aj: SFR $\approx 0.06$~M$_{\odot}$ yr$^{-1}$ \citep{pian06}.

\section{Light and colour curves of SN 2003jd}
\label{parcomparison}

\begin{figure*}
\centering
   \psfig{figure=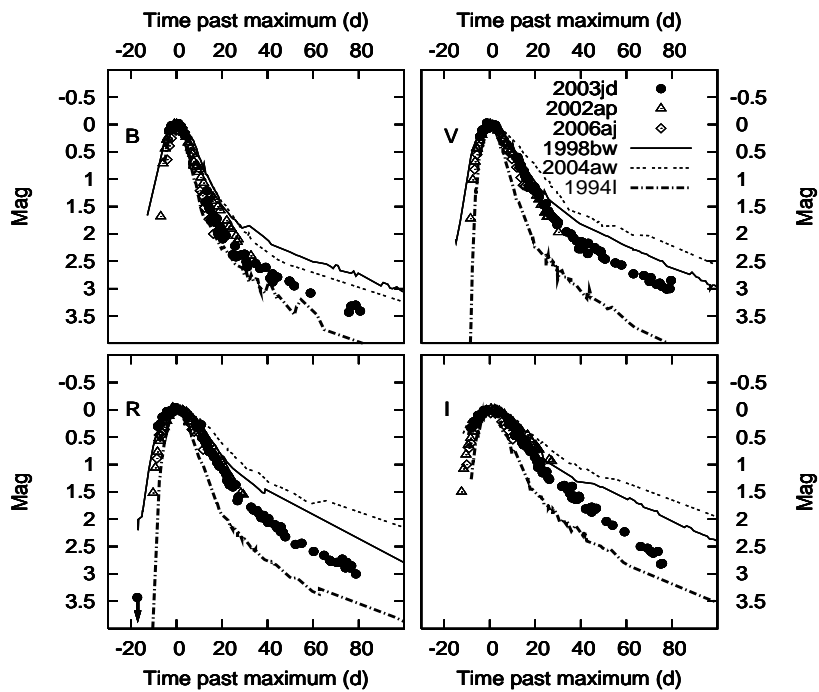,width=11.cm,height=9cm}
  \caption{Comparison  among  light   curves   of   different SNe~Ic.
  The magnitude and the phase have been shifted to coincide at 
  maximum brightness in each band.}
  \label{figconfcurve2}
\end{figure*}

In Fig. \ref{figconfcurve2} the $B$, $V$, $R$, and $I$ light curves of
SN 2003jd  are compared  with those of  a sample of  the best-observed
SNe~Ic spanning a range of observed properties, from the \emph{normal}
SN  1994I to  the  very energetic  SN  1998bw, including  intermediate
objects like SNe 2002ap, 2006aj, and 2004aw. For comparison, the light
curves are normalized to their maximum brightness in each band.

The light curves of SN 2003jd  agree well with those of SNe 2002ap and
2006aj, especially in  the $V$ and $R$ bands.  SN~1998bw and SN~2004aw
appear to be  much slower than SN 2003jd, while  SN~1994I is the other
extreme.

\begin{figure}
   \psfig{figure=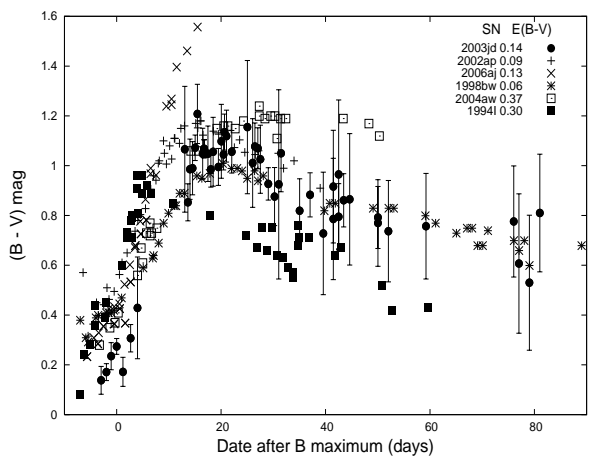,width=8cm,height=6cm,angle=-90}
  \caption{Comparison of the intrinsic colour   curves  of   different
SNe~Ic.  While among SNe~Ia the colour at  $\sim$30~d is quite
similar, SNe~Ic show a large scatter. We used the $E(B-V)$ value 
from Tab. \ref{tabparam2}}
  \label{colorcurve}
\end{figure}

From a  comparison of the $B-V$ colour  curves (Fig. \ref{colorcurve})
it appears  that the colour at  maximum, $B-V = 0.2$  mag, is actually
$\sim$0.1--0.2 mag  bluer than  that of all  other events.   SN 2003jd
evolves rapidly in colour reaching a redder colour $B-V = 1.05$ mag in
about  20~d, which  is between  SN 1998bw  ($B-V =  1.0$ mag)  and SNe
2002ap and  2004aw ($B-V  = 1.15$ mag).   SN 1994I evolves  rapidly in
color  but remains  always  bluer  ($B-V =  0.9$  mag at  $\sim$10~d),
whereas SN  2006aj reaches a  much redder colour  ($B-V = 1.6$  mag at
$\sim$15~d).

After the  peak, the colour turns  slowly back to  the blue, following
the  evolution of  SNe  1998bw  and 2002ap,  while  SN 2004aw  remains
red. This behavior could be real  or simply due to an underestimate of
the reddening of SN 2004aw,  as suggested by \cite{benetti06}.  For SN
2004aw   we    have   adopted   the   colour    excess   computed   by
\cite{taubenberger06},  $E(B-V) =  0.37$ mag,  using  the conservative
relation between reddening and the EW of the interstellar \NaID~ lines
\citep{turatto03}.

\subsection{Light-curve parameters}

In  an attempt  to  characterize quantitatively  the  light curves  of
SNe~Ic, by analogy with SNe~Ia  we use four observable parameters: (i)
the  absolute   magnitude  at  maximum  brightness   ($M$);  (ii)  the
``stretch''  factor ($s$) relative  to the  light curves  of SN~1998bw
which  is taken  as  reference\footnote{\cite{perlmutter97} introduced
the quantity $s$ for SNe~Ia; it is used to expand or contract linearly
the  time scale  of  a light  curve  in order  to  match a  template.}
\citep{perlmutter97,bloom02};   (iii)  the   difference   in  $V$-band
magnitude between  maximum and +60~d,  after expanding the  time scale
using the stretch factor ($\Delta m_{60}s$); and (iv) the slope of the
linear late-time tail ($\gamma$).

\begin{figure}
   \psfig{figure=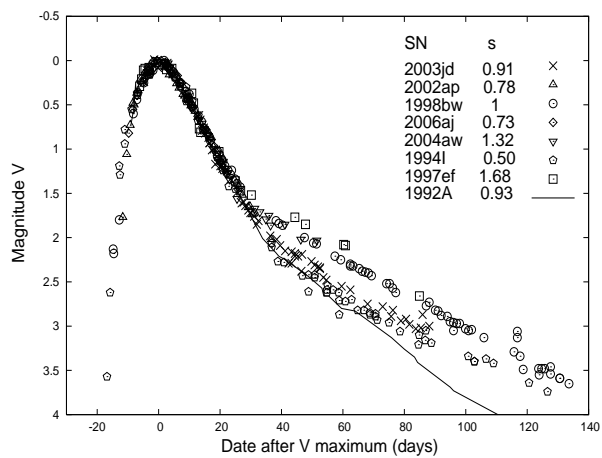,width=8cm,height=6cm}
  \caption{Comparison of $V$-band  light curves  of different SNe~Ic
  corrected for  stretch  factor. SN 1998bw is  the
  reference object with stretch factor defined to equal 1. The 
  light curves are scaled in magnitude and phase to the $V$ maximum}
  \label{figstreach}
\end{figure}

 As  shown in  Fig.~ \ref{figstreach},  the stretch  factor  allows an
excellent description of  the light curves of different  objects up to
$\sim$30~d  after  maximum,  whereas  some  spread  remains  at  later
phases.  The parameter $\Delta  m_{60}s$ has  been introduced  to take
this spread into account.  In  Fig. \ref{figstreach}, we also show the
$V$-band light  curves of  SN 1997ef  and SN 1992A.  SN 1997ef  is the
SN~Ic with  the slowest  known light curve,  even slower than  that of
SN~Ic 2004aw,  while SN 1992A  is a SN~Ia.  Up to +30~d  the stretched
$V$-band light curve of SN 1992A  follows the same shape, but later it
has  a  steeper tail  than  SNe~Ic. This  is  due  to the  inefficient
trapping  of $\gamma$-rays in  SNe~Ia than  in SNe~Ic.  At wavelengths
longer than the  $V$ band (not shown here),  the light-curve shapes of
SNe~Ia and  SNe~Ic are  distinguished by the  second peak  that SNe~Ia
show at $\sim$2 weeks after $B$ maximum.

Whether a correlation between stretch and luminosity applies to SNe Ic
as  it   does  for  SN   Ia  should  be  investigated.    However  the
uncertainties on  the reddening estimate  and the small number  of SNe
Ib/c with  good data coverage\footnote{to estimate  the stretch factor
some photometric data in the pre-maximum phase are necessary} make the
issue complex \citep{valenti08}.

The values of various parameters for SN~2003jd and other SNe~Ic are 
reported in Tab. \ref{tabparam2}.

Our analysis confirms that SN~2003jd is an intermediate case as far as
luminosity evolution  is concerned, similar to SNe  2002ap and 2006aj;
the  most extreme objects  are SN  1997ef, which  appears to  have the
slowest light curves, and SN 1994I, which is the fastest.

\section{Bolometric light curve of SN 2003jd}
\label{parametribolo}

The bolometric light curve  is a fundamental observational description
of  a  supernova,  which  can   be  readily  used  to  validate  model
predictions.  Unfortunately, because  of the restricted spectral range
covered by  the observations,  in general it  is impossible  to derive
true bolometric luminosities  for SNe. Moreover, the light  curve in a
given photometric band can be heavily affected by the evolution of few
individual lines and does not  reflect the bolometric evolution of the
SN. As a first-order approximation to the bolometric luminosity, it is
possible  to  integrate  the   SN  emission  in  the  spectral  window
accessible  from   the  ground,  redwards  of   the  ultraviolet  (UV)
atmospheric  cutoff to  the  near-IR  ($JHK$). This  is  known as  the
\emph{uvoir} bolometric luminosity.

In the  case of  SN~2003jd we have  to face the  additional limitation
that observations are available only  for the $BVRI$ optical bands. To
cope with  this limitation,  we adopt the  following strategy:  (i) we
integrate the $BVRI$ emission of SN  2003jd as a function of time (the
``quasi-bolometric light  curve''); (ii)  we compare the  results with
the same  quantity derived for other  SNe~Ic; and (iii)  we attempt to
estimate  the expected  contribution from  the unavailable  UV  and IR
bands.

To derive the $BVRI$ integrated SN luminosity, the observed magnitudes
were corrected  for reddening,  converted to the  flux density  at the
effective   wavelength   using   the   definition  of   the   standard
Johnson-Cousins  system \citep{burer,bessell83}, and  integrated using
Simpson's  rule.   Adopting  a  distance  modulus  for  each  SN,  the
integrated flux was then converted to luminosity.

\begin{figure}
   \psfig{figure=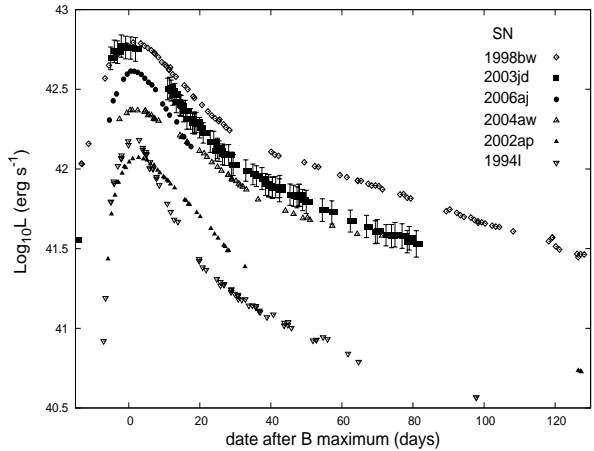,width=8cm,height=6cm}
  \caption{Quasi-bolometric ($BVRI$) light curves  of SN 2003jd and
  other SNe~Ic. We assume $H_{0} = 72$ \kms\ Mpc$^{-1}$. Extinctions and 
  distance moduli are given in Tab. \ref{tabparam2}. }
   \label{figbolocurve2}
\end{figure}

The   comparison  between   quasi-bolometric  ($BVRI$)   light  curves
(Fig. \ref{figbolocurve2})  confirms that as far  as light-curve shape
is  concerned, SN 2003jd  is most  similar to  SNe 2006aj  and 2002ap,
though  the latter  is significantly  less luminous.   Even  though SN
1998bw is  the most similar to  SN 2003jd in  absolute brightness, the
light  curve  evolves  more  slowly  than  that  of  SN  2003jd.   The
similarity  of SN~2003jd with  SN 2006aj/XRF  060218 extends  to other
properties, as we discuss in the following sections.

\begin{figure}
   \psfig{figure=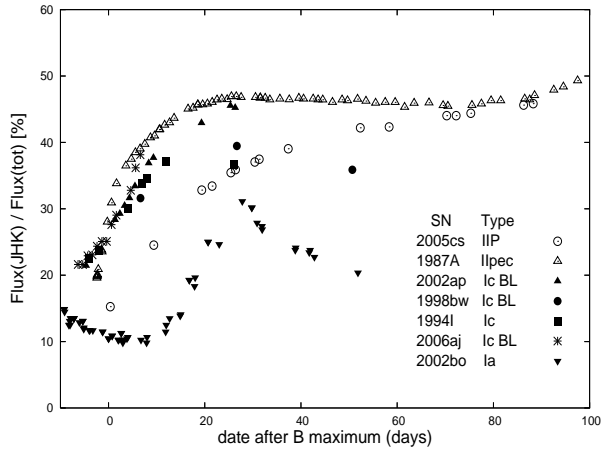,width=8cm,height=6cm}
  \caption{IR contribution to the total flux: SNe~II 
 (open  symbols), and SNe~Ia and SNe~Ic (filled  symbols).   The
 IR and optical fluxes of SN 1994I are  measured from  synthetic 
 spectra \citep{sauer06}, The optical and IR fluxes of
 the  other  SNe  are  computed integrating  the  photospheric  data,
 ({\it uvoir} for the total flux, and {\it uvoir} minus $UBVRI$ for 
 the IR flux): SN  2005cs  \citep{pastorello06,pastorello07},    
 SN 2002ap  \citep{foley03,yoshii03}, SN 1998bw \citep{galama98,patat01},
  and SN 2002bo \citep{benetti05,krisciunas04}.}
   \label{figcontributi}
\end{figure}

To estimate the contribution of the missing UV and IR spectral regions
to the bolometric luminosity, we must resort to comparisons with other
SNe.  For this purpose we  computed the contribution of $JHK$ emission
to the  total flux for different  SNe~Ic and, for  comparison, for two
SNe~II and for a SN~Ia (Fig.~\ref{figcontributi}). The IR contribution
is indeed significant, increasing from 20\% near maximum brightness up
to 40--50\%  one month later.  This is similar to  other core-collapse
SNe, and larger than in SNe~Ia, where it reaches $\sim$ 30\% one month
past $B$ maximum and decreases to 20\% at late phases. Conversely, the
UV  contribution  for  SNe~Ib/c   and  SNe~Ia  is  not  so  important:
$\sim$15\% during the first  week after explosion, decreasing later to
only 5--10\%.

Because of the similarity with  the quasi-bolometric light curve of SN
2002ap\footnote{Even  though the  light-curve  shape of  SN 2003jd  is
similar  to that  of both  SN  2006aj and  SN 2002ap,  the $JHK$  data
coverage of  SN 2002ap is larger  than that of SN  2006aj.}, we assume
that the  fractional contributions  of the UV  and IR emission  to the
bolometric  light  curve  of  SN   2003jd  are  the  same  as  for  SN
2002ap\footnote{Since the  stretch factors of  SN 2002ap ($s  = 0.78$)
and SN 2003jd ($s = 0.91$) are similar, the UV and IR contributions to
the bolometric light curve of SN 2003jd are included without using the
stretch factor as a temporal stretch.}.

We  added this  contribution  to the  quasi-bolometric ($BVRI$)  light
curve and we  used the derived bolometric light curve  of SN 2003jd to
estimate the  peak bolometric luminosity,  the slope of  the late-time
tail  and  moreover  to  compute  a rough  estimate  of  the  physical
parameters:    $M_{ej}$,    $M_{Ni}$,     and    $E_{k}$    (see    \S
\ref{parphyparam}).

The  peak bolometric  magnitude of  SN 2003jd  ($M_{bol} =  -18.65 \pm
0.20$ mag) is nearly as bright as that of SN 1998bw ($M_{bol} = -18.76
\pm 0.11$ mag), but the light curve of SN 2003jd declines more steeply
after maximum.  The slope of the  late-time tail (from 70 to 400~d) is
$\gamma = 0.015  \pm 0.001$ mag~d$^{-1}$.  This is  similar to that of
SN 1998bw  and implies a progressively increasing  transparency to the
$\gamma$-rays  produced  in  the  $^{56}$Co~\to~$^{56}$Fe  radioactive
decay.

\begin{table*}
 \centering
\caption{Parameters of the optical light curves of SNe~Ic$^a$}
\label{tabparam2}
\begin{minipage}{180mm}
\begin{tabular}{ccccccccc}
\hline
SN     & stretch     &   Magnitude & Magnitude & Magnitude & Magnitude  & $\gamma_{V}$ & $\Delta m_{60}s$ $V$ & References\\
       & $V$  &  $B$  &  $V$  &  $R$  &  $I$ & (mag d$^{-1}$) & (mag) & for the data$^b$\\
\hline
\hline
1994I  & 0.50 $\langle$0.02$\rangle$ &  $-$17.1 $\langle$0.3$\rangle$ & $-$17.5 $\langle$0.3$\rangle$ & $-$17.7 $\langle$0.2$\rangle$ & $-$17.54 $\langle$0.12$\rangle$ & 0.0184 $\langle$0.0005$\rangle$ &  2.67 $\langle$0.04$\rangle$ & R96\\
2006aj & 0.73 $\langle$0.02$\rangle$ &  $-$18.15 $\langle$0.10$\rangle$ & $-$18.67 $\langle$0.08$\rangle$ & $-$18.81 $\langle$0.06$\rangle$ & $-$18.96 $\langle$0.05$\rangle$ & -- &  -- & S06;P06;M06\\%K06 \\
2002ap & 0.78 $\langle$0.01$\rangle$ &  $-$16.79 $\langle$0.07$\rangle$ & $-$17.37 $\langle$0.05$\rangle$ & $-$17.41 $\langle$0.04$\rangle$ & $-$17.33 $\langle$0.04$\rangle$ & 0.0203 $\langle$0.0003$\rangle$ &  2.5 $\langle$0.2$\rangle$ & Y03;F03;P03\\
       &             &                &               &               &               & 0.015  $\langle$0.002$\rangle^c$ &     &        \\
2003jd & 0.91 $\langle$0.09$\rangle$ &  $-$18.7 $\langle$0.3$\rangle$ & $-$18.9 $\langle$0.3$\rangle$ & $-$19.0 $\langle$0.2$\rangle$ & $-$19.1 $\langle$0.2$\rangle$ & 0.0189 $\langle$0.0005$\rangle$ &  2.53 $\langle$0.13$\rangle$ & This work\\
1998bw & 1           &  $-$18.65 $\langle$0.07$\rangle$ & $-$19.12 $\langle$0.05$\rangle$ & $-$19.14 $\langle$0.04$\rangle$ & $-$19.04 $\langle$0.04$\rangle$ & 0.0181 $\langle$0.0004$\rangle$ &  2.23 $\langle$0.05$\rangle$ & G98;P01\\
2004aw & 1.32 $\langle$0.13$\rangle$ &  $-$17.7 $\langle$0.4$\rangle$ & $-$18.0 $\langle$0.3$\rangle$ & $-$18.1 $\langle$0.3$\rangle$ & $-$18.2 $\langle$0.2$\rangle$ & 0.017 $\langle$0.006$\rangle$ &  -- & T06\\
1997ef & 1.68 $\langle$0.10$\rangle$ &  --            & $-$17.1 $\langle$0.2$\rangle$ & --            & --            & 0.013 $\langle$0.002$\rangle$ &  2.02 $\langle$0.13$\rangle$  & M04\\
\hline
\end{tabular}

$^a$The  extinctions, the distance  moduli, and the data used for the 
SNe~Ic are as follows. 
SN 2002ap: $E(B-V)=0.09$ mag \citep{yoshii03}, $\mu = 29.46$ mag \citep{sharina96};
SN 2004aw: $E(B-V)=0.37$ mag, $\mu  =  34.17$ mag \citep{taubenberger06};
SN 1998bw: $E(B-V)=0.06$ mag, $\mu  =  32.76$ mag \citep{patat01};
SN 1994I: $E(B-V)=0.3$ mag, $\mu   =  29.60$ mag \citep{sauer06};
SN 2006aj: $E(B-V)=0.13$ mag, $\mu   =  35.6$ mag \citep{pian06};
SN 1997ef: $E(B-V)=0.04$ mag, $\mu   =  33.5$ mag  \citep{iwamoto00}; and
SN 2003jd: $E(B-V)=0.14$ mag,  $\mu   =  34.46$ mag (this work).

$^b$R96 = \cite{richmond96}; S06 = \cite{sollerman06}; P06 = \cite{pian06}; M06 = \cite{mirabal06}; Y03 = \cite{yoshii03}; F03 = \cite{foley03}; P03 = \cite{pandey03}; P01 = \cite{patat01}; G98 = \cite{galama98}; T06 = \cite{taubenberger06}; I00 = \cite{iwamoto00}; M04 = \cite{mazzali04}.

$^c$SN 2002ap shows an inflection point at $t \approx 250$~d. The slope 
is measured in two different temporal ranges: 70--250~d and 250--400~d.  

\end{minipage}
\end{table*}

\subsection{Spectral evolution and stretch factor}
\label{paragrafostretch}

In \S  \ref{parcomparison} we introduced  the stretch factor,  to take
into account the different time  evolution of SN~Ic light curves.  The
spectral  evolution seems  to follow  a similar  behavior:  SN 1997ef,
which  has  the  slowest  light  curve  in  our  sample,  still  shows
photospheric  features  at 150~d  \citep{mazzali04},  while SN  1994I,
which has  the fastest  light curve, starts  to show  nebular features
just two months after maximum brightness.

In  order to investigate  whether this  behavior is  a signature  of a
spectral  sequence, we compared  spectra of  SNe~Ic using  a stretched
time scale.  In  practice, we took as reference  the inferred epoch of
explosion for  each event (see Tab.  \ref{tabparam3}),  and we labeled
spectra with a  ``stretched" phase obtained by scaling  the phase past
explosion by the stretch factor derived from the light-curve shape.
 
\begin{figure}
   \psfig{figure=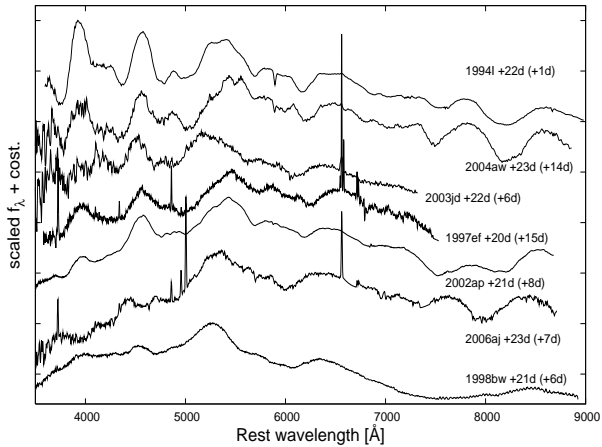,width=8cm,height=6cm,angle=-90}
  \caption{Comparison of spectra of SNe~Ic at a stretched time of
$\sim$3 weeks past explosion (computing time in SN 1998bw frame). 
Next to each object name, we report the phase after the 
explosion in the stretched time scale, and, in parentheses, 
the phase relative to the observed $B$ maximum.}
  \label{figcomp1}     
\end{figure}

\begin{figure}
   \psfig{figure=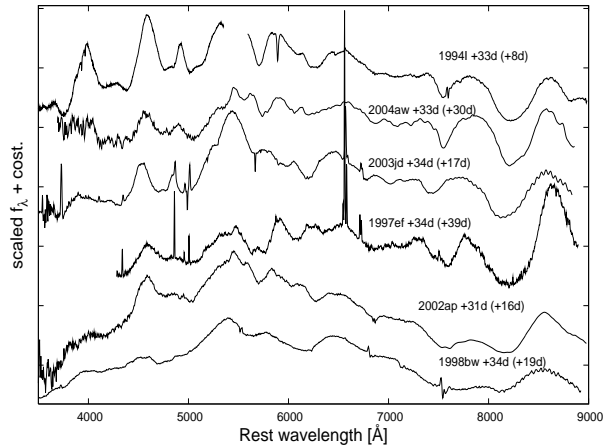,width=8cm,height=6cm,angle=-90}
  \caption{Comparison of spectra of SNe~Ic at a stretched-time of
  $\sim$5 weeks past explosion (computing time in SN 1998bw frame).
Next to each object name, we report the phase after the 
explosion in the stretched time scale and, in parentheses, 
the phase relative to the observed $B$ maximum.}  
The spectrum of SN  2006aj is missing, because the object
was too close to the Sun at the relevant time.
  \label{figcomp2}     
\end{figure}

We then  compared spectra of  different objects at the  same stretched
time.   This is  shown  in Figs.   \ref{figcomp1} and  \ref{figcomp2},
where the epochs  are $\sim$3 weeks and $\sim$5  weeks past explosion,
respectively, in the time frame SN~1998bw.

As the  figures show,  with this  approach part of  the spread  of the
sample is removed.  SN 1994I, with the fastest light curves (SN 1994I,
$s=0.50$)  and SN  2004aw, with  the second-slowest  light  curves (SN
2004aw, $s=1.32$), appear very similar both in their line profiles and
expansion velocities.  The major  differences between the two SNe seen
close to  5700--5800~\AA~ in  Fig. \ref{figcomp3}, disappears  in Fig.
\ref{figcomp1}.     Indeed,   the    strong   absorption    close   to
5700--5800~\AA\  in SN~1994I  at  +2 weeks  (mainly  caused by  \NaID)
develops at about +1 month in SN~2004aw (see Fig. \ref{figcomp2}).

The  stretch  of  the  time  scale  does  not  completely  remove  the
heterogeneity of  the sample.  Indeed, SNe 2002ap,  2006aj, and 2003jd
show broad  lines at all epochs, but  never as broad as  in SN 1998bw.
The  features of  SN 1994I  and  SN 2004aw  are the  narrowest of  the
sample.  While  the time stretching  for the light curves  works well,
for spectroscopy  it can partially  help in the comparison  of similar
objects (SN 2004aw and SN 1994I).

Other parameters  are required  to explain the  remaining differences.
The  different line  widths are  probably connected  to  the expansion
velocity   and  to  the   kinetic  energy   per  ejected   mass  ratio
($E_{k}/M_{ej}$).  Moreover, the  asymmetric explosion and the viewing
angle make the scenario even more  complex, as is already the case for
nebular spectra \citep{maeda06a} and light curves \citep{maeda06b}

The  flux deficiency  in the  blue  part of  the spectra  seen in  the
figures for  a few  SNe~Ic deserves some  discussion. \cite{mazzali02}
suggested that  the blue deficiency  is related to the  high expansion
velocity and  to the  consequent severe blending  of the  many 
\FeII lines at these wavelengths.

The fact that  SN 2003jd has higher blue emission  than SNe 1998bw and
2006aj, and a relatively narrow Fe feature at $\sim$4900~\AA, strongly
suggests that  the Fe  layers have slower  expansion velocities  in SN
2003jd  than in  other objects.  In a  scenario where  iron  is mainly
expelled along  the jet \citep{maeda02}, the  lower expansion velocity
of the  Fe lines in SN  2003jd is in  agreement with the idea  that in
this SN, the jet was out of the line of sight \citep{mazzali05}.

\begin{figure}
   \psfig{figure=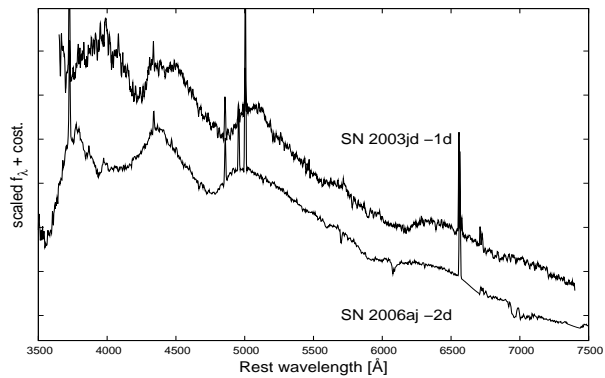,width=8cm,height=5cm,angle=-90}
  \caption{Comparison of spectra of SN 2003jd and SN 2006aj close 
   to $B$ maximum. }
   \label{pippo2}     
\end{figure}

However, except  the blue excess,  the object with spectra  similar to
those   of   SN   2003jd   is   SN  2006aj.   This   is   visible   in
Figs.  \ref{figcomp3}  and  \ref{figcomp1},  and it  is  also  evident
comparing   the    spectra   close   to    maximum   brightness   (see
Fig. \ref{pippo2}).

\section{Was there a GRB associated with  SN~2003jd?}
\label{parmissinggrb}

The close  similarity of SN~2003jd with  SN~2006aj/XRF 060218 prompted
the quest for a possible GRB associated with SN 2003jd.

The main  argument against  a relativistic jet  along or off  from the
line of sight is the  non-detection of SN 2003jd at radio wavelengths,
despite the  extensive search for  radio emission extending  from soon
after explosion to 1.6~yr  later \citep{soderberg06b}. The early radio
observations are sensitive to both typical GRBs and sub-energetic GRBs
(e.g., XRF-SN 2006aj) viewed along  the jet axis, while at late times,
since the  ejecta are roughly spherical, radio  observations should be
sensitive to both typical GRBs and sub-energetic GRBs viewed off-axis.

For SN  2003jd, the earliest radio  upper limit (at  $t \approx 10$~d)
implies  a  radio   luminosity  a  factor  of  10   below  that  of  a
sub-energetic   GRB   radio    afterglow   such   as   XRF-SN   2006aj
\citep{soderberg06b},  at a  comparable epoch.   Therefore,  radio and
X-ray  observations  show  no  evidence  for  any  typical  GRB  or  a
sub-energetic GRB viewed along the axis.

On  the other  hand, the  double-peaked [O\,{\sc  i}] emission  in the
nebular  spectra  suggests  significant  asymmetry of  the  explosion,
consistent  with the presence  of a  jet-like structure  oriented away
from the line of sight. Even though there was no late-time radio of SN
2003jd,  suggesting no  evidence  for a  typical  GRB viewed  off-axis
\citep{soderberg06b}, an  off-axis GRB in  a low-density circumstellar
medium cannot be excluded.

We investigated the  (lack of) evidence for a  possible GRB associated
to  SN2003jd  \citep{gcn2434,gcn2439}.   Assuming  that  the  GRB  was
simultaneous with the  SN, the temporal window to  be searched is well
defined \citep[$16 \pm 2$ Oct. 2003; ][]{sauer07}.

To search  for a possible GRB  associated with SN 2003jd,  we used the
data of  the interplanetary  network (IPN).  In  the 14--18  Oct. 2003
search window, the IPN consisted  of the HETE and RHESSI spacecraft in
low-Earth orbit, INTEGRAL (SPI-ACS detector) at 0.5 light-seconds from
Earth,  Wind (Konus detector)  at 1.8  light-seconds from  Earth, Mars
Odyssey (High Energy Neutron Detector and Gamma Sensor Head detectors)
at 268 light-seconds from Earth,  and Ulysses (GRB experiment) at 2940
light-seconds  from  Earth.  All  instruments  were  on and  operating
nominally throughout  the search window, although Konus-Wind  was in a
solar particle event, which raised its threshold slightly.

In this configuration, the IPN acts as a true all-sky monitor for GRBs
and other fast transients, with virtually no Earth-occultation or duty
cycle considerations.  Its lower-energy threshold is $\sim$25 keV, and
its flux and fluence thresholds,  although very dependent on the burst
time history, are $\sim$1  photon $cm^{-2}$ $s^{-1}$ and $10^{-6}$ erg
$cm^{-2}$,  respectively.    At  these  levels,   the  probability  of
detecting a burst is roughly 50\%.  Bursts with fluxes and fluences an
order  of magnitude  below  these  values can  be  detected, but  with
probabilities  of 30--40\%.   The IPN  has observed  and  detected the
SN-associated GRBs 980425 (SN  1998bw), 021211 (SN 2002lt), 030329 (SN
2003dh), and  031203 (SN  2003lw).  It did  not detect XRF  060218 (SN
2006aj), probably because the spectrum of this event was too soft.

Three GRBs were clearly detected in the search window by more than one
spacecraft.   All of  them can  therefore be  classified  as confirmed
cosmic  events, and  could  be localized.   However, the  localization
regions  excluded the  position of  SN  2003jd in  all cases.   Other,
weaker  events  may  also  have  been  present,  particularly  in  the
Konus-Wind data, but their  interpretation is complicated by the solar
particle background.

Thus,  if SN  2003jd  indeed produced  a  GRB, we  are  left with  the
following possible conclusions:  (i) the GRB was beamed  away from us;
(ii) the GRB  was weak, below the IPN flux  and fluence thresholds; or
(iii) the GRB had a soft spectrum, with little or no emission above 25
keV.  Any combination of the above properties is also possible.

\section{Physical parameters from the bolometric light curve}
\label{parphyparam}

The modelling of  the bolometric light curve can  be used to constrain
the parameters of the explosion, in particular $M_{ej}$, $M_{Ni}$, and
$E_{k}$.  For this purpose, the SN luminosity evolution can be divided
into  two  phases  according  to  the  different  dominating  physical
processes.

\begin{itemize}  
\item \emph{Photospheric  Phase} ($t \le  30$~d past  explosion): We
adopt the simple analytical  model developed by \cite{arnett82} for
SNe~Ia, which  is well  suited  also for  core-collapse SNe lacking a
hydrogen  recombination phase (SNe~Ib/c).  The  model  assumes homologous
expansion  of the ejecta,  spherical symmetry,  and that all radioactive Ni
is located  in  the  centre.    It  adopts  a  constant  optical  opacity
$k_{opt}$, a small initial radius before explosion ($R_{0}$~\to~0), and
the diffusion approximation for photons (i.e., the ejecta  are optically
thick; see  App.  \ref{modello}).  With respect  to \cite{arnett82}, we
include also the energy produced  by the $^{56}$Co~\to~$^{56}$Fe decay,
not only the energy produced by the $^{56}$Ni~\to~$^{56}$Co decay.

\item \emph{Nebular Phase}  ($t \ge 60$~d past  explosion): At late
phases, with the  decrease of  the optical  depth of  the  ejecta, the
emitted luminosity is directly  related to the instantaneous amount of
energy deposition by the $\gamma$-rays  from the $^{56}$Co decay, by 
the  $\gamma$-rays coming from the electron-positron annihilation,
and by the kinetic energy of the positrons 
\citep[cf.][]{sutherland84,cappellaro97}.
\end{itemize}  

In the  transition phase ($30 \le  t \le 60$~d), where  the ejecta are
not   completely  thick   but   also  not   sufficiently  thin,   both
approximations  fail. This  portion  of the  light  curve was  ignored
during the fit.

The  dominant physical effect  to consider  is that  in SNe~Ic,  as in
SNe~Ia, the  ejecta become rapidly and  progressively more transparent
to  $\gamma$-rays,  which  thus  escape.   This is  accounted  for  by
multiplying  the decay energy  by an  absorption probability  (for the
$\gamma$-ray  photons) and  by  an annihilation  probability (for  the
positrons).   In   the  assumption  of   homologous  expansion,  these
probabilities decline with time.

\begin{figure}
   \psfig{figure=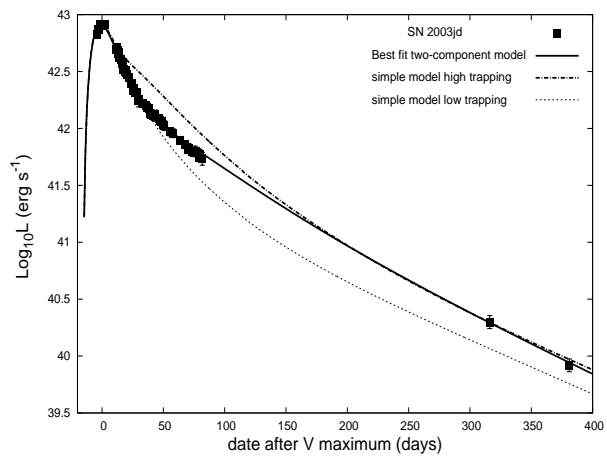,width=8cm,height=6cm}
  \caption{Light-curve fit with  simple models (low and high $\gamma$-ray 
 trapping; dotted and dot-dashed lines, respectively) and with a 
 two-component model (solid line) of the bolometric light curve of 
 SN 2003jd. While the simple models cannot fit the photospheric 
 phase and the nebular phase in a consistent way, the two-component model 
account for the light curve in both phase.}
  \label{figmodel03jd}     
\end{figure}

While this simple model gives good fits to the overall light curves of
SNe~Ia  \citep[see appendix  \ref{modello}]{arnett82},  for SNe~Ic  we
found an inconsistency between the parameters derived from the fitting
of the  early and  late-time light curves.   In particular,  the rapid
evolution of the light curve of the photospheric phase requires a high
transparency  of $\gamma$-rays  (see Fig.   \ref{figmodel03jd}) which,
however, results in too low a predicted luminosity at late phase.

This was already noted by  \cite{maeda03}, who obtained a fair overall
fit to  the light  curves of SNe~Ic  by schematically dividing  the SN
ejecta into two regions: a high-density inner region and a low-density
outer region with high and low $\gamma$-ray trapping, respectively.

It is fair to assume that  the luminosities emitted by the two regions
sum  up to produce  the total  luminosity when  the emission  from the
inner  region is  not absorbed  by the  outer ejecta.  This  is indeed
reasonable in  the nebular  phase, when the  outer layer  is basically
transparent.  Conversely, during the  photospheric phase  the opposite
regime holds:  the inner  region is optically  thick and  the emerging
luminosity is  a small fraction  of the total luminosity.   Indeed, we
found that  at the epoch of  maximum, for typical  fit parameters, the
flux emerging from  the inner region is at most  15--30\% of the total
flux.

Using this two-component model increases the number of parameters from
3  to  6, which  were  finally reduced  to  5  assuming a  homogeneous
distribution of  Ni in the inner  and outer layers.   The 5 parameters
are $M_{Ni}(tot)$,  $M_{ej}(tot)$, $E_{k}(inner)$, $E_{k}(outer)$, and
the fraction of the total mass concentrated in the dense, low-velocity
inner ejecta.

In  addition, the  width of  the light  curve during  the photospheric
phase  ($w_{lc}$) and its  slope in  the nebular  phase ($\gamma_{lc}$)
have a similar dependence on $M_{ej}$ and $E_{k}$:

\begin{equation}
\label{eqdeg}
w_{lc} \propto \frac{M^{3/4}_{ej}}{E^{1/4}_{k}}~~~~{\rm and} \\
\gamma_{lc} \propto  \frac{M_{ej}}{E^{1/2}_{k}}.
\end{equation}
%while $L_{\gamma}$ depends on the ratio $M_{ej}/E^{1/2}_{k}$

\noindent
This introduces a  degeneracy in the estimate of  $M_{ej}$ and $E_{k}$
which  can  be  removed  by  using  some  additional  constraint  from
spectroscopy.  More  precisely, assuming a homogeneous  density of the
ejecta, the photospheric velocity at maximum is related to the kinetic
energy and the ejected mass through the relation \citep{arnett82}

\begin{equation}
v_{ph} \approx \frac{3}{5}~~\frac{2 E_{k}}{M_{ej}.}
\end{equation}

As  we mentioned  above, at  maximum the  outer ejecta  give  the main
contribution to the observed luminosity. Hence, assuming a homogeneous
density  in this region  and measuring  the photospheric  velocity, we
constrain the kinetic energy of the outer layers for a given mass, and
thus resolve the degeneracy of Eq. \ref{eqdeg}.

For  the  photospheric  velocity  we  adopt the  values  derived  from
spectral  modelling (see  Tab~7). Uncertainties  are typically  of the
order of $10\%$ .

The best  fit of the bolometric light  curve of SN 2003jd  is shown in
Fig.~\ref{figmodel03jd}  (thick  solid   line).   The  values  of  the
physical  parameters  derived  from   the  model  are  $M_{Ni}(tot)  =
0.36 \pm 0.04$~M$_{\odot}$,            $M_{ej}(tot)           =
3.0_{-0.5}^{+0.5}$~M$_{\odot}$,  and $E_{k}(tot) =  7_{-2}^{+3} \times
E_{51}$.  The uncertainty in the nickel  mass is mainly due to that in
the  distance. The  uncertainty in  the photospheric  velocity affects
both the uncertainty in $M_{ej}$ and $E_{k}$.

The values of $M_{ej}$, $M_{Ni}$,  and $E_{k}$ obtained by fitting the
light   curves   of   our   sample   of   SNe~Ic   are   reported   in
Tab.\ref{tabparam3}.  For comparison,  we give  estimates of  the same
parameters  obtained  from  the  spectral  and  light-curve  modelling
available in  the literature. While  the $^{56}$Ni and  $E_{K}$ values
are consistent  with those in  literature, the values of  $M_{ej}$ are
smaller of $\sim30~ \%$. A similar underestimate of the ejected mass,
together with an underestimate of the kinetic energy,
was found by \cite{richardson06} using an analytical light curve model
similar to our simple model. The two-components model reproduce better 
the light curve giving also a better estimate of the kinetic energy.
Therefore we considered, to be conservative,
an  error of  30$\%$ in  our estimate  of the  ejected mass  of SN 2003jd
($M_{\rm ej} = 3 \pm 1$ M$_{\odot}$).

\begin{table*}
 \centering
\begin{minipage}{180mm}
\caption{Physical parameters for SNe~Ic derived from models of the
bolometric light curves$^a$}  
\label{tabparam3}
%\begin{tabular}{rl}
\begin{tabular}{@{}ccccccccccc@{}}
\hline
SN  & $V_{ph}^a$ & $T_{exp}^b$ &$M_{Ni}(tot)$ & $M_{ej}(tot)$ & $E_{k}$(tot)  & $E_{k}$(inner) & $M_{Ni}(inner)/$ & $M_{Ni}(*)$ & $M_{ej}(*)$ & $E_{k}(*)$\\
       & \kms\       & days             & M$_{\odot}$ & M$_{\odot}$& $10^{51}$ \ene\ & $10^{51}$ \ene\ &$M_{Ni}(tot)$ & M$_{\odot}$ & M$_{\odot}$ & $10^{51}$ \ene      \\
\hline		    
\hline
1994I & 11000 &11 $\pm$ 2&0.065$\pm$0.03& $0.7_{-0.1}^{+0.3}$&$0.9_{-0.1}^{+0.7}$&$0.0017_{-0.001}^{+0.005}$&$0.22_{-0.04}^{+0.03}$&$\sim$\emph{0.07}&$\sim$\emph{0.9}&$\sim$\emph{1}\\
2002ap & 14000  &  8 $\pm$ 1 &   0.073$\pm$0.02 & $1.6_{-0.1}^{+0.5}$& $2.4_{-0.4}^{+2}$& $0.05_{-0.01}^{+0.08}$&0.40-0.56&$\sim$\emph{0.1}&\emph{$\sim$2.5}&$\sim$\emph{4}\\
2004aw & 14000  &  $\sim$ 15  &   0.21$\pm$0.03  & $5_{-1}^{+2}$       & $8_{-1.5}^{+8}$  & $0.18_{-0.13}^{+0.4}$ & 0.38-0.50 &\emph{--} &\emph{--}&\emph{--}\\ 
2003jd & 13500  &  13 $\pm$ 2 &   0.36$\pm$0.04  & 3.0$\pm$0.5          & $7_{-2}^{+3}$    & $0.02_{-0.01}^{+0.05}$&  0.18-0.28 & -- & -- & --  \\
1998bw & 18000  &  15 $\pm$ 1 &  0.49$\pm$0.04 & $8_{-0.1}^{+0.5}$& $34_{-1}^{+3} $ & $0.26_{-0.01}^{+0.04}$& $0.29_{-0.01}^{+0.01}$&\emph{0.5} &\emph{10--11} & \emph{30--60}\\
%2006aj & 17000  &  9.5 $\pm$ 1   &               &                  &       &          &            &           \\
\hline
\end{tabular} 
\\  

$^a$The $E(B-V)$ values and the distance moduli  are
reported in Tab. \ref{tabparam2}. For comparison, we report in the last
three columns [labeled with (*)] the values of $M_{ej}$, $M_{Ni}$, and
$E_{k}$, computed  with the spectral and light-curve  modelling: SN
1998bw and SN 1994I  \citep{nomoto01}, SN 2002ap \citep{mazzali07b}, and SN
2004aw \citep{kawabata07}.

$^b V_{ph}$ is the photospheric velocity  close to the  
maximum of the  bolometric light curve.

$^c T_{exp}$ is  the time elapsed from the explosion to the $B$  
maximum. The $V_{ph}$  and $T_{exp}$ values are  taken  from  
spectral modelling: SN 1998bw \citep{mazzali01}, SN 1994I
\citep{sauer06}, SN 2002ap \citep{mazzali07b}, SN 2006aj \citep{mazzali06a},
SN 2004aw \citep{kawabata07}, and SN 2003jd \citep{sauer07}.  
\end{minipage}
\end{table*}

This estimate  of the ejected  mass indicates that the  progenitor C+O
star had  a mass of $\sim$3--7~M$_{\odot}$, which  evolved most likely
from a main-sequence star with $M_{\rm ms}$~$\sim$22--28~M$_{\odot}$.

If the star  was on the low side  of this range, it is  likely to have
collapsed  to   a  neutron  star.    The  small  remnant   mass  would
than favour the eject ion of  a large mass of $^{56}$Ni as in
SN 2006aj,  possibly also via a magnetar.  However, the ejecta
of SN 2006aj  were relatively spherically symmetric \citep{mazzali07},
unlike  those of  SN 2003jd.  Although asphericity  may be  present in
most,  or  possibly  all  core-collapse  SNe, it  is  not  clear  that
low-energy events can  be as highly aspherical as  either SN 2003jd or
SN   1998bw  \citep{mazzali01,maeda02}.   On   the  other   hand,  the
morphological  similarity between  the  very aspherical  ejecta of  SN
2003jd  and those  of SN  1998bw \citep{mazzali05}  is an  argument in
favour of the collapsar scenario for  SN 2003jd as well, in which case
the mass of the progenitor would  probably have been close to the high
side of the range mentioned above.

A  large  $E_{k}/M_{ej}$ ratio  is  required  to  reproduce the  broad
features.   Indeed, with  $E_{51}/M_{ej}  \approx 2.3$,  SN 2003jd  is
second only to SN 1998bw ($E_{51}/M_{ej} \approx 4.2$).  SN 2004aw has
energy  comparable to that  of SN  2003jd, but  a larger  ejected mass
($E_{51}/M_{ej}  \approx 1.6$).   This is  the reason  for  the slower
evolution of  the light  curve, accompanied by  narrow lines as  in SN
1994I ($E_{51}/M_{ej} \approx 1.3$).

In conclusion, SN 2003jd is  a broad-line SN~Ic, more massive and more
energetic  than the broad-line  SNe 2006aj  and 2002ap,  but certainly
less massive and energetic than SN 1998bw.

\section{Summary}

We  have presented  optical photometry  and spectroscopy  of SN~2003jd
spanning  from  3~d  before  $B$-band  maximum  to  $\sim$400~d  after
maximum. SN 2003jd shows the typical spectral features of a SN~Ic, but
broadened as in the case of SNe 2002ap and 2006aj.

The  light curves  are similar  in shape  to those  of SNe  2002ap and
2006aj,  but the  peak  luminosity is  rather  similar to  that of  SN
1998bw.

The  comparison with  a  sample of  well-studied  SNe~Ic confirms  the
heterogeneity  of this  SN  class. However,  the  application of  time
stretching (as  for SNe~Ia) helps  in homogenizing this class  of SNe,
reducing  the  differences among  the  light  curves and  facilitating
spectral  comparisons.  Different  explosion  energies and  progenitor
masses are likely  to be the main reasons  for the discrepancies.  The
different viewing  angles from which  the SNe are observed  (given the
evidence  for asymmetric  explosions)  may also  explain  some of  the
differences, but this  effect is probably not large,  at least for the
light-curve  shape  \citep{maeda06b}.  It  may,  however,  affect  the
spectra \citep{tanaka07}.   The $E_{k}/M_{ej}$  ratio seems to  be the
main factor in producing broad features.

The similarity to  SN 2006aj, the presence of  broad features, and the
asymmetry shown by  the oxygen double peak in  the nebular spectra are
in favour  of an asymmetric explosion  oriented away from  the line of
site.   The  radio  observations  argue  that the  asymmetry  did  not
extended to relativistic velocities, although uncertainties in the CSM
mean that 2003jd could still be a mis-directed GRB.

Finally,  we used a  simple model  for the  bolometric light  curve to
obtain the main  physical parameters of SN 2003jd.  The derived values
confirm that SN 2003jd is a  broad-line SN similar to SN 2002ap and SN
2006aj,  but   with  a  larger   ejected  mass  ($M_{ej}  =   3.0  \pm
1.0$~M$_\odot$) and  kinetic energy ($E_{k}(tot)  = 7_{-2}^{+3} \times
E_{51}$ erg)  and producing a  large quantity of $^{56}Ni$  ($M_{Ni} =
0.36 \pm  0.04$~M$_\odot$).  Comparing  the physical parameters  of SN
2003jd with those of other SNe~Ic \citep{nomoto07}, SN 2003jd appears
to represent  a link between  broad-lined SNe (2002ap and  2006aj) and
GRB-associated supernovae (SNe 1997ef, 1998bw, 2003dh, and 2003lw).

\section*{Acknowledgments}
We thank D. Branch to have refereed the paper, 
S.  Taubenberger,  F.  Saitta,  and  G.  Bono  for  helpful
suggestions.   We   are   grateful   to  V.   Pal'shin,   E.   Mazets,
S. Golenetskii, I.  Mitrofanov, A.  Sanin, W. Boynton, A. von Kienlin,
G.  Lichti,  A.  Rau,  G.   Ricker,  D. Lamb,  and  J.-L.  Atteia  for
contributing  Konus, Mars  Odyssey,  INTEGRAL, and  HETE  data to  the
interplanetary  network  to   search  for  an  accompanying  gamma-ray
burst. A.V.F. thanks the Lorentz  Center in Leiden for its hospitality
during the workshop ``From Massive Stars to Supernova Remnants,'' when
this paper was finalized.

This work was supported by grant  n. 2006022731 of the PRIN of Italian
Ministry  of University  and Sci.  Reaserch.  A.  Gal-Yam acknowledges
partial   support   by    NASA   through   Hubble   Fellowship   grant
HST-HF-01158.01-A awarded  by STScI, which is operated  by AURA, Inc.,
for  NASA,  under  contract   NAS  5-26555.   A.V.F.'s  group  at  the
University of  California, Berkeley  is supported by  National Science
Foundation   (NSF)    grant   AST--0607485   and    by   the   TABASGO
Foundation.  KAIT was  made possible  by generous  donations  from Sun
Microsystems,    Inc.,   the   Hewlett-Packard    Company,   AutoScope
Corporation, Lick Observatory, the  NSF, the University of California,
and the Sylvia  \& Jim Katzman Foundation.  K.  Hurley is grateful for
IPN  support under  the  following grants  and  contracts: JPL  958056
(Ulysses),  MIT SC-A-293291  (HETE),  NAG5-13080 (RHESSI),  NAG5-12614
(INTEGRAL), and NAG5-11451 (HETE and Mars Odyssey).

This  paper   is  based  on  observations  made   with  the  following
facilities:  the  European  Southern  Observatory  telescopes  (Chile)
obtained   from  the   ESO/ST-ECF  Science   Archive   Facility  (Prog
ID.  074.D-0161A),  the 10~m  Keck~I  telescope  at  the W.   M.  Keck
Observatory   (Hawaii),  the  robotic   1.5~M  telescope   at  Palomar
Observatory (California),  the Italian National  Telescope Galileo (La
Palma),  the  1.82~m Copernico  telescope  of  the Asiago  Observatory
(Italy), the 2.3~m Advanced  Technology Telescope at the Siding Spring
Observatory  (Australia), the  1~m telescope  at the  Wise Observatory
(Israel),  the  Subaru  8.2~m  telescope of  the  National  Astronomic
Observatory of  Japan (Hawaii),  KAIT and the  Shane 3~m  telescope at
Lick  Observatory (California), and  the 1.5~m  telescope at  the Fred
Lawrence Whipple Observatory (Arizona).  We are grateful to the staffs
at  all of  the  telescopes for  their  assistance.  The  W. M.   Keck
Observatory  is  operated  as   a  scientific  partnership  among  the
California Institute of Technology,  the University of California, and
the National Aeronautics and Space Administration; the Observatory was
made  possible by the  generous financial  support of  the W.  M. Keck
Foundation.

%\bibliography{articolo}

\begin{thebibliography}{99}
\bibitem[\protect\citeauthoryear{Alard}{2000}]{alard2000}  Alard,  C.\
2000, A\&AS, 144, 363
\bibitem[\protect\citeauthoryear{Arnett} {1982}]{arnett82} 
Arnett, W.~D.\ 1982, ApJ, 253, 785 
\bibitem[\protect\citeauthoryear{Benetti et al.}{2005}]{benetti05} 
Benetti, S., et al.\ 2005, ApJ, 623, 1011
\bibitem[\protect\citeauthoryear{Benetti et al.}{2006}]{benetti06} 
Benetti, S., Cappellaro, E., Turatto, M., Taubenberger, S., 
Harutyunyan, A., \& Valenti, S.\ 2006, ApJ, 653, L129
\bibitem[\protect\citeauthoryear{Bessell}{1983}]{bessell83} 
Bessell, M.~S.\ 1983, PASP, 95, 480 
\bibitem[\protect\citeauthoryear{Bloom et al.}{2002}]{bloom02} 
Bloom, J.~S., et al.\ 2002, ApJL, 572, L45 
\bibitem[\protect\citeauthoryear{Branch    et    al.}{2006}]{branch06}
Branch, D.,  Jeffery, D.~J., Young,  T.~R., \& Baron, E.\  2006, PASP,
118, 791
\bibitem[\protect\citeauthoryear{Burket et al}{2003}]{burket03} 
Burket, J., Swift, B., \& Li, W. 2003, IAU Circular 8232
\bibitem[\protect\citeauthoryear{Buser \& Kurucz} {1978}]{burer} 
Buser, R., \& Kurucz, R.~L.\ 1978, A\&A, 70, 555
\bibitem[\protect\citeauthoryear{Campana et al.}{2006}]{campana06} 
Campana, S., et al.\ 2006, Nature, 442, 1008 
\bibitem[\protect\citeauthoryear{Cappellaro et
al.}{1997}]{cappellaro97} Cappellaro, E., Mazzali, P.~A., Benetti, S.,
Danziger, I.~J.,  Turatto, M., Della Valle, M.,  \& Patat,  F.\ 1997,
A\&A, 328, 203
\bibitem[\protect\citeauthoryear{Clocchiatti                         et
al.}{1996}]{clocchiatti96}    Clocchiatti,    A.,   Wheeler,    J.~C.,
Brotherton,  M.~S.,  Cochran,  A.~L.,  Wills, D.,  Barker,  E.~S.,  \&
Turatto, M.\ 1996, ApJ, 462, 462
\bibitem[\protect\citeauthoryear{Clocchiatti \& Wheeler}{1997}] 
 {clocchiatti97}  Clocchiatti,  A.,  \& Wheeler,  J.~C.\ 1997, ApJ, 491, 375
\bibitem[\protect\citeauthoryear{Colgate   et  al}  {1997}]{colgate97}
Colgate  S.   A.,  Fryer,  C.  L., \&  Hand,  K.  P.  1997,  in
Thermonuclear Supernovae, ed. P. Ruiz-Lapuente,   R.  Canal, \&  
J.   Isern  (Dordrecht: Kluwer), p. 273
\bibitem[\protect\citeauthoryear{Elmhamdi  et  al.}{2006}]{elmhamdi06}
Elmhamdi, A., Danziger, I.~J., Branch,  D., Leibundgut, B., Baron, E.,
\& Kirshner, R.~P. 2006, \aap, 450, 305
\bibitem[\protect\citeauthoryear{Filippenko}{1982}]{filippenko82}
Filippenko, A.~V.\ 1982, PASP, 94, 715
\bibitem[\protect\citeauthoryear{Filippenko}{1997}]{filippenko97}
Filippenko, A.~V.\ 1997, ARAA, 35, 309
\bibitem[\protect\citeauthoryear{Filippenko                          et
al.}{2001}]{filippenko01} Filippenko, A. V., Li, W. D., Treffers, R. R.,
  \& Modjaz, M. 2001, in Small-Telescope Astronomy on Global Scales, 
   ed. W. P. Chen, C. Lemme, \& B. Paczy\'{n}ski (San Francisco: 
   ASP), p. 121
\bibitem[\protect\citeauthoryear{Filippenko                          et
al.}{2003}]{filippenko03} Filippenko, A. V., Foley, R. J., \& Swift, B.
2003, IAU Circular 8234
\bibitem[\protect\citeauthoryear{Filippenko}{2005}]{filippenko05}
Filippenko, A.~V.\ 2005, in The Fate of the Most Massive
    Stars, ed. R. Humphreys \& K. Stanek (San Francisco: ASP), p. 33
\bibitem[\protect\citeauthoryear{Foley et al.}{2003}]{foley03} 
Foley, R.~J., et al.\ 2003, PASP, 115, 1220 
\bibitem[\protect\citeauthoryear{Freedman et al.}{2001}]{freedman01} 
Freedman, W.~L., et al.\ 2001, ApJ, 553, 47 
\bibitem[\protect\citeauthoryear{Galama et al.}{1998}]{galama98} 
Galama, T.~J., et al.\ 1998, Nature, 395, 670 
\bibitem[\protect\citeauthoryear{Gal-Yam    et   al.}{2002}]{galyam02}
Gal-Yam, A., Ofek, E.~O., \& Shemmer, O.\ 2002, MNRAS, 332, L73
\bibitem[\protect\citeauthoryear{Hjorth    et    al.}{2003}]{hjorth03}
Hjorth, J., et al.\ 2003, Nature, 423, 847
\bibitem[\protect\citeauthoryear{Hurley et al.}{2003}]{gcn2439} 
Hurley, K., et al.\ 2003, GRB Coordinates Network, 2439, 1 
\bibitem[\protect\citeauthoryear{Iwamoto et al.}{1998}]{iwamoto98} 
Iwamoto, K., et al.\ 1998, Nature, 395, 672 
\bibitem[\protect\citeauthoryear{Iwamoto et al.}{2000}]{iwamoto00} 
Iwamoto, K., et al.\ 2000, ApJ, 534, 660 
\bibitem[\protect\citeauthoryear{Kawabata et al.}{in prep.}]{kawabata07} 
Kawabata, K., 2007, in preparation
\bibitem[\protect\citeauthoryear{Kennicutt}{1998}]{kennicutt98}
Kennicutt, R.~C., Jr.\ 1998, ARA\&A, 36, 189
\bibitem[\protect\citeauthoryear{Krisciunas et al.}{2004}]{krisciunas04} 
Krisciunas, K., et  al.\ 2004, AJ, 128, 3034
\bibitem[\protect\citeauthoryear{Landolt}{1992}]{landolt92}    Landolt,
A.~U. 1992, AJ, 104, 340
%\bibitem[MacFadyen \& Woosley(1999)]{macfadyen99} MacFadyen, A.~I., 
%\& Woosley, S.~E.\ 1999, ApJ, 524, 262 
\bibitem[\protect\citeauthoryear{Maeda et al.}{2002}]{maeda02} Maeda,
K., Nakamura, T., Nomoto, K., Mazzali, P.~A., Patat, F., \& Hachisu,
I.\ 2002, ApJ, 565, 405
\bibitem[\protect\citeauthoryear{Maeda  et al.}{2003}]{maeda03} Maeda,
K., Mazzali, P.~A.,  Deng, J., Nomoto, K., Yoshii,  Y., Tomita, H., \&
Kobayashi, Y.\ 2003, ApJ, 593, 931
\bibitem[\protect\citeauthoryear{Maeda et al.}{2006a}]{maeda06a} 
Maeda, K., Nomoto, K., Mazzali, P.~A., \& Deng, J.\ 2006, ApJ, 640, 854 
\bibitem[\protect\citeauthoryear{Maeda et al.}{2006b}]{maeda06b} 
Maeda, K., Mazzali, P.~A., \& Nomoto, K.\ 2006, ApJ, 645, 1331
\bibitem[\protect\citeauthoryear{Maeda et al.}{2007}]{maeda07} 
Maeda, K., et al.\ 2007, ApJ, 658, L5
\bibitem[\protect\citeauthoryear{Maeda et al.}{submitted}]{maeda07b} 
Maeda, K., et al. submitted
\bibitem[\protect\citeauthoryear{Malesani et al.}{2004}]{malesani04} 
Malesani, D., et al.\ 2004, ApJL, 609, L5 
\bibitem[\protect\citeauthoryear{Mazzali   et   al.}{2001}]{mazzali01}
Mazzali, P.~A., Nomoto,  K., Patat, F., \&  Maeda, K.\ 2001, ApJ, 559,
1047
\bibitem[\protect\citeauthoryear{Mazzali  et    al.}{2002}]{mazzali02}
Mazzali, P.~A., et al.\ 2002, ApJ, 572, L61
\bibitem[\protect\citeauthoryear{Mazzali et al.}{2004}]{mazzali04} 
Mazzali, P.~A., Deng, J., Maeda, K., Nomoto, K., Filippenko, A.~V., 
\& Matheson, T.\ 2004, ApJ, 614, 858 
\bibitem[\protect\citeauthoryear{Mazzali  et    al.} {2005}]{mazzali05}
Mazzali, P. A., et al, 2005, Science, 308, 1284
\bibitem[\protect\citeauthoryear{Mazzali  et   al.}{2006}]{mazzali06a}
Mazzali, P.~A., et al.\ 2006, Nature, 442, 1018
\bibitem[\protect\citeauthoryear{Mazzali et al.}{2007}]{mazzali07} 
Mazzali, P.~A., et al.\ 2007, ApJ, 661, 892 
\bibitem[\protect\citeauthoryear{Mazzali et al.}{2007}]{mazzali07b} 
Mazzali, P.~A., et al.\ 2007, ArXiv e-prints, 708, arXiv:0708.0966 
\bibitem[\protect\citeauthoryear{Mirabal et al.}{2006}]{mirabal06} 
Mirabal, N., Halpern, J.~P., An, D., Thorstensen, J.~R., \& Terndrup, 
D.~M.\ 2006, ApJ, 643, L99 
\bibitem[\protect\citeauthoryear{Modjaz et al.}{2007}]{modjaz07} 
Modjaz, M., Kewley, L., Kirshner, R.~P., Stanek, K.~Z., 
Challis, P., Garnavich, P.~M., Greene, J.~E., \& Prieto, J.~L.\ 2007, 
ArXiv Astrophysics e-prints, arXiv:astro-ph/0701246
\bibitem[\protect\citeauthoryear{Nomoto    et    al.}{2001}]{nomoto01}
Nomoto,  K.,  Mazzali, P.~A.,  Nakamura,  T.,  Iwamoto, K.,  Danziger,
I.~J.,  \&  Patat, F., ``Supernovae  and  Gamma-Ray Bursts:  the
Greatest Explosions since the Big Bang'' ed. Mario Livio, Nino Panagia, 
Kailash Sahu in Cambridge University Press, ISBN 0-521-79141-3, 2001, 
p. 144 - 170
%\bibitem[\protect\citeauthoryear{Nomoto   et     al.}{2000}]{nomoto00}
%Nomoto  K. ,Maeda  K., Nakamura T.,   Iwamoto K., Suzuki, T.,  Mazzali
%P. A., Turatto M., Danziger, I. J., Patat, F.  2000, AIPC, 526, 622N
\bibitem[\protect\citeauthoryear{Nomoto et al.}{2007}]{nomoto07} 
Nomoto, K., Tominaga, N., Tanaka, M., Maeda, K., Suzuki, T., 
Deng, J.~S., \& Mazzali, P.~A.\ 2007, 
ArXiv Astrophysics e-prints, arXiv:astro-ph/0702472
\bibitem[\protect\citeauthoryear{Ofek et al.}{2003}]{gcn2434} 
Ofek, E.~O., Poznanski, D., Gal-Yam, A., \& Lipkin, Y.\ 2003, 
GRB Coordinates Network, 2434, 1 
\bibitem[\protect\citeauthoryear{Pagel  et al.}{1979}]{pagel79} Pagel,
B.~E.~J., Edmunds, M.~G., Blackwell, D.~E., Chun, M.~S., \& Smith, G.\
1979, MNRAS, 189, 95
\bibitem[\protect\citeauthoryear{Pandey et al.}{2003}]{pandey03} 
Pandey, S.~B., Anupama, G.~C., Sagar, R., Bhattacharya, D., Sahu, D.~K., 
\& Pandey, J.~C.\ 2003, MNRAS, 340, 375 
\bibitem[\protect\citeauthoryear{Pastorello et al.}{2006}]{pastorello06} 
Pastorello, A., et al.\ 2006, MNRAS, 370, 1752 
\bibitem[\protect\citeauthoryear{}{Pastorello et al. in prep.}]{pastorello07} 
Pastorello, A., et al.\ , in preparation
\bibitem[\protect\citeauthoryear{Patat }{1996}]{patat06} 
Patat, F. 1996, PhD thesis, University of Padova
\bibitem[\protect\citeauthoryear{Patat  et al.}{2001}]{patat01} Patat,
F., et al. 2001, ApJ, 555, 900
\bibitem[\protect\citeauthoryear{Perlmutter et al.}{1997}]{perlmutter97} 
Perlmutter, S., et al.\ 1997, ApJ, 483, 565 
\bibitem[\protect\citeauthoryear{Pettini  \&  Pagel}{2004}]{pettini04}
Pettini, M., \& Pagel, B.~E.~J.\ 2004, MNRAS, 348, L59
\bibitem[\protect\citeauthoryear{Pian et al.}{2006}]{pian06}  Pian,
E., et al.\ 2006, Nature, 442, 1011
\bibitem[\protect\citeauthoryear{Prieto et al.}{2007}]{prieto07} 
Prieto, J.~L., Stanek, K.~Z., \& Beacom, J.~F.\ 2007, ArXiv e-prints, 
707, arXiv:0707.0690 
\bibitem[\protect\citeauthoryear{Richmond et al.}{1996}]{richmond96} 
Richmond, M.~W., et al.\ 1996, AJ, 111, 327
\bibitem[Richardson et al.(2006)]{richardson06} Richardson, D., 
Branch, D., \& Baron, E.\ 2006, AJ, 131, 2233
\bibitem[\protect\citeauthoryear{Sauer  et al.}{2006}]{sauer06} Sauer,
D.~N.,   Mazzali, P.~A., Deng, J.,  Valenti,  S.,   Nomoto,   K., \&
Filippenko, A.~V.\ 2006, MNRAS, 369, 1939
\bibitem[\protect\citeauthoryear{Sauer  et  al.}{in prep.}]{sauer07}
Sauer, D. N., et al., in preparation
\bibitem[\protect\citeauthoryear{Schlegel et al.}{1998}]{schlegel98} 
Schlegel, D.~J., Finkbeiner, D.~P., \& Davis, M.\ 1998, ApJ, 500, 525 
\bibitem[\protect\citeauthoryear{Sharina  et  al.}{1996}]  {sharina96}
Sharina, M.~E., Karachentsev,  I.~D., \& Tikhonov,  N.~A.\ 1996, A\&AS,
119, 499
\bibitem[\protect\citeauthoryear{Soderberg et al.}{2006}]{soderberg06a} 
Soderberg, A.~M., et al.\ 2006, Nature, 442, 1014 
\bibitem[Soderberg et al.(2006)]{soderberg06b} Soderberg, A.~M., 
Nakar, E., Berger, E., \& Kulkarni, S.~R.\ 2006, ApJ, 638, 930 
\bibitem[\protect\citeauthoryear{Sollerman et al.}{2006}]{sollerman06}
Sollerman, J., et al.\ 2006, A\&A, 454, 503
\bibitem[\protect\citeauthoryear{Spergel et al.}{2003}]{spergel03} 
Spergel, D.~N., et al.\ 2003, ApJS, 148, 175
\bibitem[\protect\citeauthoryear{Stanek et al.}{2003}]{stanek03} 
Stanek, K.~Z., et al.\ 2003, ApJ, 591, L17 
\bibitem[\protect\citeauthoryear{Sutherland et al.}{1984}]{sutherland84} 
Sutherland, P. G., \& Wheeler, J. C. 1984, ApJ, 280, 282
\bibitem[\protect\citeauthoryear{Tanaka et al.}{2007}]{tanaka07} 
Tanaka, M., Maeda, K., Mazzali, P.~A., \& Nomoto, K.\ 2007, 
ArXiv e-prints, 708, arXiv:0708.3242 
\bibitem[\protect\citeauthoryear{Taubenberger et al.}{2006}]{taubenberger06} 
Taubenberger, S., et al.\ 2006, MNRAS, 874
\bibitem[\protect\citeauthoryear{Thomas et al.}{2004}]{thomas04} 
Thomas, R.~C., Branch, D., Baron, E., Nomoto, K., Li, W., \& 
Filippenko, A.~V.\ 2004, ApJ, 601, 1019 
\bibitem[\protect\citeauthoryear{Turatto    et  al.} {2003}]{turatto03}
Turatto, M., Benetti, S., \& Cappellaro, E. 2003, in From Twlight to 
Highlight: The   Physics    of  Supernovae, ed. W.  Hillebrandt \& B. 
Leibundgut (Berlin: Springer), p. 200
\bibitem[\protect\citeauthoryear{Valenti          et          al.}{in
prep.}]{valenti08} Valenti, S. 2008, in preparation
\bibitem[\protect\citeauthoryear{van     den     Bergh     et     al.}
{2005}]{vandenbergh05} van  den Bergh, S.,  Li, W., \& Filippenko, A. V. 2005,
PASP, 117, 773 
\bibitem[\protect\citeauthoryear{Yoshii  et  al.}  {2003}]  {yoshii03}
Yoshii, Y., et al. 2003, ApJ, 592, 467\\
\end{thebibliography}

\appendix
\section[]{Fitting the bolometric light curves of SNe~Ia}
\label{modello}
Following  the prescriptions of  \cite{arnett82} for  the photospheric
phase and those of \cite{sutherland84} and \cite{cappellaro97} for the
nebular phase, a  simple model was adopted for  fitting the bolometric
light    curves   of    Type   I    supernovae   (SNe    lacking   the
hydrogen-recombination phase):

\begin{itemize}  
\item  \emph{Photospheric phase}  ($t  \le 30$~d  past explosion):  We
assume  homologous expansion  for the  ejecta, spherical  symmetry, no
mixing  for  Ni, constant  optical  opacity  $k_{opt}$, small  initial
radius   before  explosion   ($R_{0}$  \to   0),  and   the  diffusion
approximation for photons (optically thick ejecta).  These assumptions
are mainly those proposed by \cite{arnett82} for SNe~Ia, with the main
exception that  both $^{56}$Ni and $^{56}$Co are  considered as energy
sources.   With these  assumptions,  the luminosity  evolution in  the
photospheric phase is described  by the following equation, a function
of $E_{k}$, $M_{Ni}$, and $M_{ej}$:
\end{itemize}

\begin{eqnarray}
\label{earlyphase}
 \lefteqn{  L_{ph}(t) = M_{Ni}  e^{-x^{2}} \times } \nonumber  \\ &
&\bigg[ ( {\epsilon}_{Ni}-{\epsilon}_{Co} ) \int_{0}^{x} A(z) dz
+ {\epsilon}_{Co} \int_{0}^{x} B(z) dz \bigg],
\end{eqnarray}

with
\begin{equation}
A(z)= 2 z e^{-2zy+z^{2}}~~~{\rm and}\\
B(z)= 2 z e^{-2zy+2zs+z^{2}},\\
\end{equation}
where $x \equiv t/\tau_{m}$,  $y \equiv \tau_{m}/(2\tau_{Ni})$, and $s
\equiv  (\tau_{m}(\tau_{Co}-\tau{Ni})/(2\tau_{Co}\tau_{Ni})$. The
energies produced  in one second  by one gram  of $^{56}$Ni and $^{56}$Co are,
respectively, $\epsilon_{Ni}  =  3.90 \times 10^{10}$ 
erg~s$^{-1}$~g$^{-1}$  and $\epsilon_{Co} = 6.78 \times 10^{9}$ 
erg~s$^{-1}$~g$^{-1}$ \citep{sutherland84,cappellaro97}.

Let $\tau_{Ni}$ and $\tau_{Co}$ be the decay time of $^{56}$Ni and 
$^{56}$Co, respectively, while $\tau_{m}$ is the time scale  
of the light curve. With the assumption of  
a homogeneous density, $\tau_{m}$ is
\begin{equation}
\label{eqvcsle}
\tau_{m}= \bigg( \frac{k_{opt}}{\beta c} \bigg)^{1/2}\bigg( \frac{10 M^{3}_{ej}}{3 E_{k}} \bigg)^{1/4}, \\ 
\end{equation}
where $\beta \approx 13.8$ is a constant of integration
\citep{arnett82} and $k_{opt}$ is the optical opacity.

\begin{itemize}  
\item \emph{Nebular phase} ($t \ge 60$~d past explosion): During the
nebular phase, the light curve  is determined by the energy
deposition of the $^{56}$Ni~\to~$^{56}$Co~\to~$^{56}$Fe radioactive 
decay chain.  The light curve is described by the following
equation \citep{cappellaro97,sutherland84}:
\end{itemize}  
\begin{equation}
L_{neb}(t)=S^{Ni}(\gamma)+S^{Co}(\gamma)+S^{Co}_{e^{+}}(\gamma)+S^{Co}_{e^{+}}(KE),\\
\end{equation}
where $S^{Ni}(\gamma)$  is    the source energy  of    the  nickel decay,
\begin{equation}
S^{Ni}(\gamma)=    M_{Ni}      \epsilon_{Ni}     e^{-t/\tau_{Ni}},\\
\end{equation}
while  the other terms are  the amount  of energy  deposited  by the
radioactive decay of $^{56}$Co (81\% of the energy is released as $\gamma$-rays
and 19\% as  positrons). If ${\cal E}$ is the rate  of energy production by
the $^{56}$Co decay,
\begin{equation}
 {\cal E} = M_{Ni} \epsilon_{Co} (e^{-t/\tau_{Co}}-e^{-t/\tau_{Ni}}),\\
\end{equation}
then the energy $S^{Co}(\gamma)$ generated by the deposition of $\gamma$-rays 
from $^{56}$Co decay can be written as
\begin{equation}
S^{Co}(\gamma)=0.81~ {\cal E}~ (1-e^{-(F/t)^{2}}),\\
\end{equation}
the  amount energy  deposited  by the  $\gamma$-rays  produced in  the
positron annihilation $S^{Co}_{e^{+}}(\gamma)$ can be written as
\begin{equation}
S^{Co}_{e^{+}}(\gamma)=0.164~ {\cal E}~ (1-e^{-(F/t)^{2}})(1-e^{-(G/t)^{2}}), 
\end{equation}
and the source  energy  due  to the kinetic energy of  the positrons
$S^{Co}_{e^{+}}(KE)$ can be written as
\begin{equation}
S^{Co}_{e^{+}}(KE)=0.036~ {\cal E}~ (1-e^{-(G/t)^{2}})
\end{equation}

The incomplete  trapping of $\gamma$-rays and positrons  is taken into
account      in      the      terms      $(1-e^{-(F/t)^{2}})$      and
$(1-e^{-(G/t)^{2}})$.  Under the  assumption of  homologous expansion,
the   absorption   probability  for   $\gamma$-ray   photons  can   be
approximated by $(1-e^{-(F/t)^{2}})$  and the annihilation probability
for    positrons   can    be    approximated   by    $(1-e^{-(G/t)2})$
\citep{clocchiatti97}.  The  parameters $F$  and $G$ are  functions of
the ejected mass, kinetic energy, and opacity \citep{clocchiatti97}:

\begin{equation}
F=\sqrt{\bigg(C(\rho) k_{\gamma} M_{ej}^{2}\bigg)/E_{k}},
\end{equation}

\begin{equation}
G=\sqrt{\bigg(C(\rho) k_{e^{+}} M_{ej}^{2}\bigg)/E_{k}} 
\end{equation}
where  $C(\rho)$  is  a  function  of the  density  $\rho$.   Using  a
homogeneous    density,     $k_{\gamma}=0.027$,  and    $k_{e^{+}}=7$
\citep{clocchiatti97}, $F \approx 32M_{ej}/\sqrt{E_{51}}$ and $G \approx
515M_{ej}/\sqrt{E_{51}} =  16.1F$  \citep[see also ][]{colgate97}.

\begin{figure} 
   \psfig{figure=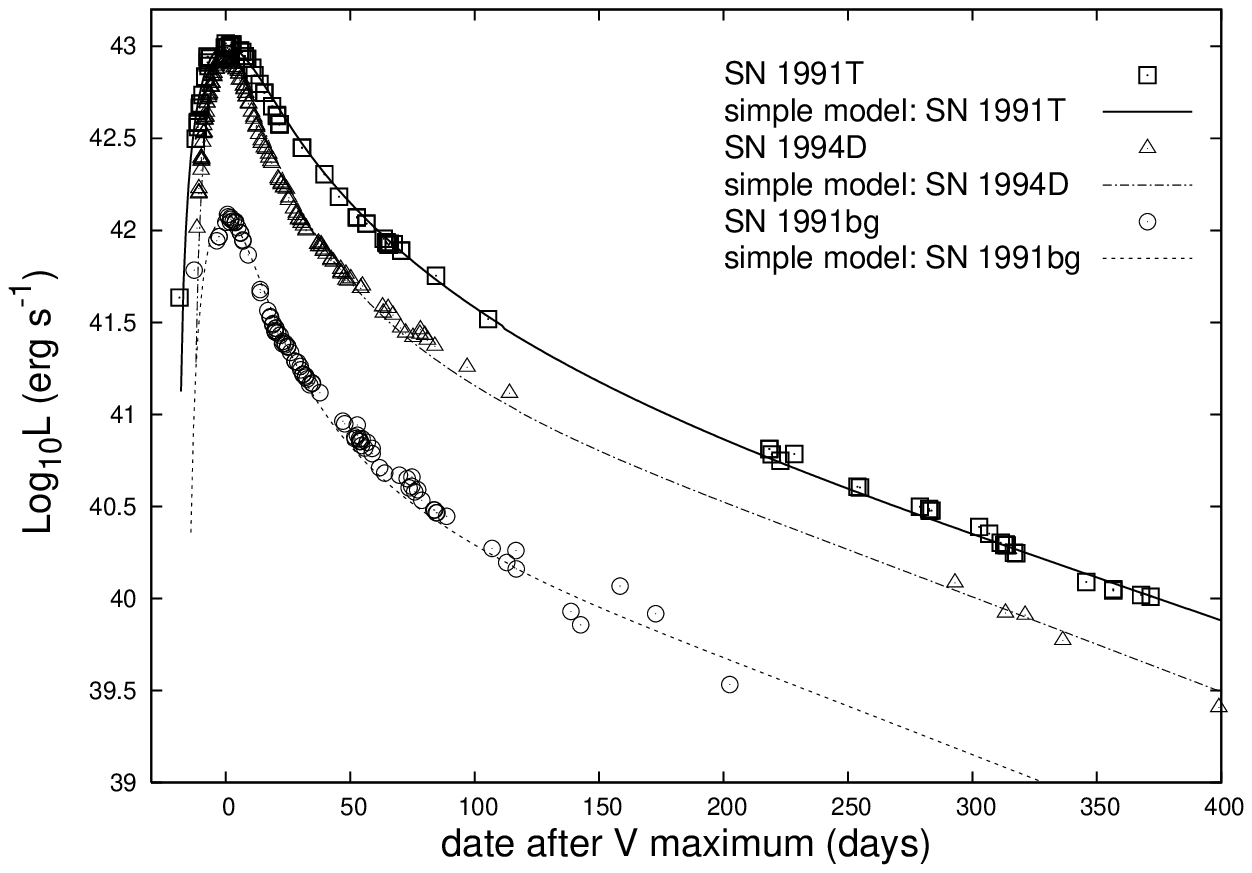,width=8cm,height=6cm}
 \caption{Fit of  the quasi-bolometric light curves of three SNe~Ia
  with the simple model.}
  \label{figsimplemodel}     
\end{figure}

Using the  information coming  from the fits  to the  photospheric and
nebular  phases, we  obtain  a  rough estimate  of  the main  physical
parameters of  the SN,  without developing a  specific model  for each
object.  This  simple model gives a  good fit to  the bolometric light
curves   of  SNe~Ia   (see  Fig.   \ref{figsimplemodel}),   while  the
two-component model  described in  \S \ref{parphyparam} is  needed for
SNe~Ic.

\end{document}